\documentclass[acmsmall]{acmart}

\AtBeginDocument{%
  }

\setcopyright{acmlicensed}
\acmJournal{PACMMOD}
\acmYear{2024} \acmVolume{2} \acmNumber{3 (SIGMOD)} \acmArticle{149} \acmMonth{6}\acmDOI{10.1145/3654952}

\acmPrice{15.00}
\acmISBN{978-1-4503-XXXX-X/18/06}

\usepackage{graphicx}  
\graphicspath{{images}}
\usepackage[a-1b]{pdfx} 
\usepackage{array}
\usepackage{newtxtext,newtxmath}
\usepackage{algorithmic}
\usepackage[ruled,vlined]{algorithm2e}
\usepackage{pgfplots,pgfplotstable}  
\usepackage{booktabs}
\usepackage{balance}
\usepackage{multirow}
\usepackage{subcaption}
\pgfplotsset{compat=newest}
\usepackage{balance}
\usepackage{ulem}
\usepackage{etoolbox}
\usepackage{setspace} 
\usepackage{pgf-pie}
\usepackage{tikz}
\usetikzlibrary{patterns}

\usepackage{xcolor}

\definecolor{teal}{HTML}{00AEB3}

\newcommand\redsout{\bgroup\markoverwith{\textcolor{red}{\rule[0.5ex]{2pt}{0.4pt}}}\ULon}
\newcommand{\edit}[1]{\textcolor{black}{#1}}

\newcommand{\eat}[1]{}
\newcommand{\sysname}{Lorentz\xspace}
\newcommand{\msft}{Microsoft\xspace}
\newcommand{\doppler}{Doppler\xspace}
\newcommand{\metadata}{profile data\xspace}

\newcommand{\pg}{Azure PostgreSQL DB\xspace}

\usepackage{enumitem}

\newcounter{enum}

\received{October 2023}
\received[revised]{January 2024}
\received[accepted]{February 2024}

\begin{document}
\title{\sysname: Learned SKU Recommendation Using Profile Data\\
  (DMDS)}


\author{Nick Glaze}
\affiliation{%
	\institution{Microsoft}
	\streetaddress{1 Memorial Drive}
	\city{Cambridge}
	\state{MA}
	\postcode{02142}
    \country{USA}
}
\email{nickglaze@microsoft.com}
\authornotemark[1]

\author{Tria McNeely}
\affiliation{%
	\institution{Microsoft}
	\streetaddress{1 Memorial Drive}
	\city{Cambridge}
	\state{MA}
	\postcode{02142}
    \country{USA}
}
\email{triamcneely@microsoft.com}
\authornote{Authors contributed equally.}

\author{Yiwen Zhu}
\affiliation{%
	\institution{Microsoft}
	\city{Mountain View}
    \country{USA}
}
\email{yiwzh@microsoft.com}
\authornotemark[1]

\author{Matthew Gleeson}
\affiliation{%
\institution{Microsoft}
	\streetaddress{1 Memorial Drive}
	\city{Cambridge}
	\state{MA}
	\postcode{02142}
    \country{USA}
}
\email{mattgleeson@microsoft.com}

\author{Helen Serr}
\affiliation{%
\institution{Microsoft}
	\streetaddress{1 Memorial Drive}
	\city{Cambridge}
	\state{MA}
	\postcode{02142}
    \country{USA}
}
\email{helenserr@microsoft.com}

\author{Rajeev Bhopi}
\affiliation{%
	\institution{Microsoft}
	\streetaddress{One Microsoft Way}
	\city{Redmond}
	\state{WA}
	\postcode{98052}
    \country{USA}
}
\email{rajbho@microsoft.com}

\author{Subru Krishnan}
\affiliation{%
	\institution{Microsoft}
	\city{Barcelona}
    \country{Spain}
}
\email{subru@microsoft.com}

\renewcommand{\shortauthors}{Nick Glaze et al.} 


\begin{abstract}

In response to diverse demands, cloud operators have significantly expanded the array of service offerings, often referred to as Stock Keeping Units (SKUs) available for computing resource configurations. Such diversity has led to increased complexity for customers to choose the appropriate SKU. In the analyzed system, only 43\% of the resource capacity was rightly chosen. Although various automated solutions have attempted to resolve this issue, they often rely on the availability of enriched data, such as workload traces, which are unavailable for newly established services. Since these services amass a substantial volume of telemetry from existing users, cloud operators can leverage this information to better understand customer needs and mitigate the risk of over- or under-provisioning. Furthermore, customer satisfaction feedback serves as a crucial resource for continuous learning and improving the recommendation mechanism.

In this paper, we present \sysname, an intelligent SKU recommender for provisioning new compute resources that circumvents the need for workload traces. \sysname leverages customer \edit{\metadata} to forecast resource capacities for new users based on detailed profiling of existing users. Furthermore, using a continuous learned feedback loop, \sysname 
tailors capacity recommendations according to customer performance vs. cost preferences captured through satisfaction signals.
Validated using the production data from provisioned VMs supporting \pg, we demonstrate that
\sysname outperforms user selections and existing defaults, reducing slack by >60\% without increasing throttling. Evaluated using synthetic data, \sysname's personalization stage iteratively learns the user preferences over time with high accuracy.

\end{abstract}



\maketitle



\section{Introduction}
\label{sec:intro}
\begin{figure}[t]
    \centering
    \resizebox{4.7cm}{1.98cm}{%
    \begin{tikzpicture}
        \pie[
            /tikz/every pin/.style={align=center},
            radius=1.3,
            text=legend,
            color={red!60, orange!60, green!60},
            rotate=90,
            explode={0, 0, 0, 0},
        ]{
            38/Under-provisioned,
            19/Over-provisioned,
            43/Right-sized}
    \end{tikzpicture}
    }
    \caption{Users improperly provision many resources on \pg (flexible server).}
    \vspace{-0.4cm}
    \label{fig:pie-chart}
\end{figure}
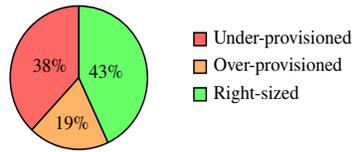

The availability of public cloud services enables easy access to a wide range of data services with diverse analytic requirements (e.g., SQL/NoSQL databases, streaming, machine learning, business insight analysis, and a lot more)~\cite{visionzhu}. However, the complexity of choosing optimal offerings increases significantly when a large number of choices are exposed. Cloud operators have invested tremendous efforts to provide ``knob-free'' services by automating numerous aspects throughout the onboarding process. For instance, various tools~\cite{kang2010cloudle, kopaneli2015model} have been developed to aid in selecting the appropriate target, facilitating smooth migration from on-premises to the cloud. However, many of these tools still necessitate substantial \edit{input} from customers. Model-driven approaches like \doppler~\cite{cahoon2022doppler} construct customer profiles and utilize clustering analysis to glean preferences from existing customers, which can then be applied to new ones, achieving full automation. Automated configuration tuning, leveraging ML methods such as Bayesian optimization, has gained recent attention as a viable approach for constant, online configuration tuning~\cite{curino2020mlos, zhu2021kea, selfadapt, selftune, selfmanage}.
However, tools designed to support such scenarios typically require a detailed historical trace for each workload they operate on, and thus are more applicable to optimizing existing workloads than new ones. As of late, there is limited discussion about automating SKU selection for initial resource provisioning. At \msft, for example, users of \pg are simply provided a static default SKU, which they select much more often than is appropriate; this results in significant COGS loss as well as performance throttling. As shown in Figure \ref{fig:pie-chart}, only 43\% of users correctly provision their services, further exacerbating their losses.

In this paper, we tackle the selection of \textit{initial} optimal SKUs for newly-created services (such as PaaS databases or VM instances). We discuss a method to circumvent the requirement for workload trace data, particularly for services whose configurations are inconvenient to modify after being initially provisioned (e.g., when autoscaling induces long reboot times). In approaching this problem, several challenges emerge:

\textbf{Scarcity of input data [C1]}: In contrast to scenarios involving service migration, where telemetry can provide comprehensive insights into each workload, initial workload configuration presents a distinct challenge due to the limited availability of information. In this scenario, inference must be made without data related to the workload traces, instead relying on customer \edit{\metadata} like industry, company, or customer IDs. 
\edit{With the recent advent of machine learning and data science, enriched \metadata has demonstrated increasing utility in facilitating detailed customer management tasks~\cite{gorodetsky2014agent, sawhney2019uber}, showcasing significant potential for enhancing SKU recommendation processes.}

\textbf{Customer heterogeneity [C2]}:
Cloud customers have widely varying preferences in the trade-off between price and performance. Configuration recommendations must account for these preferences and integrate them into the solution. Employing a one-size-fits-all solution for all customers is prone to resource over-provisioning or under-provisioning, failing to align with their distinct preferences.

\textbf{Explainability [C3]}:
Since configuration recommendations have a significant influence on future cost and performance outcomes, customers often seek a comprehensive understanding of the underlying reasoning behind them. Customers rely on this transparency to ensure they are well-informed to make final judgments about the fidelity of recommendations they are presented with. \edit{Good explainability also leads to better user adoption given comprehension of the decision-making process and trust in results, encouraging potential user interaction.}

\textbf{Lack of properly labeled data [C4]}:
We observed a considerable proportion of instances in \pg to be either over-provisioned or under-provisioned (Figure \ref{fig:pie-chart}). Existing customers' choices are therefore unreliable as ``labels'' for training any downstream configuration recommendation models.


\subsubsection*{\textbf{Introduction to \sysname}} 
To address these challenges, we propose \textit{\sysname}, an automated system for initial SKU recommendation that is applicable to provision VMs, DBs, or other cloud resources. 
\sysname consists of three stages: (1) \textit{capacity rightsizing} for exiting workloads based on slack (unused quantity of resources) and throttling (severity of resource crunching), leveraging auto-scaling strategies to account for biases in the training data; (2) an innovative explainable \textit{SKU recommender} that bases predictions on similar customers' instances; and (3) a continuous running \textit{personalizer} that dynamically learns the price-performance preference for each customer based on feedback signals such as Customer Reported Incident (CRI) information through a message propagation process. 

Specifically,

(1) \sysname aggregates data from diverse origins, including subscription (each corresponds to one user account) particulars, resource specifics, customer engagements, and feedback from current users to make each SKU recommendation [\textbf{C1}]. \sysname draws insights from the selection patterns of existing users, predicated on their similarities. The rationale behind this is that similarity between customers' compute resources, subscriptions, or \edit{\metadata} might imply similarity between their resource capacity needs. For example, databases provisioned by Coca-Cola and Pepsi might have similar needs, since they share a common industry. \sysname opens up the opportunity to gain useful business insight from \edit{\metadata} to support critical decision-making and is easily extensible should more enriched information becomes available.


(2) That \sysname recommends based on similarity in customer \edit{\metadata} enables design of explainable recommender models [\textbf{C3}]. When generating each recommendation, \sysname can additionally provide the ``search result" from the referenced database to help customers understand what characteristics of themselves or their new compute resource were useful for the recommendation. We provide two distinct algorithms to characterize customers and their services to be provisioned.

(3) \sysname learns each customer or customer segment’s price-performance preferences by assessing historical customer interactions with resource provisioning, scaling actions, and performance-related CRIs [\textbf{C2}]. We provide a comprehensive feedback loop to constantly adjust the recommendation strategy at the individual level to capture customer heterogeneity. Each customer is tagged with \textit{preference scores} which is updated periodically to reflect learning of their preference along the price-performance trade-off. We developed an innovative learning algorithm to account for potential network impact across different resources provisioned by the same user, or within the same industry (e.g., Banking or Retail).

(4) \sysname provides flexible, modular, and general solutions that can support provisioning, pooling, and rightsizing operations across a multitude of data ecosystems and beyond. The general solution is straightforwardly extensible as additional \edit{\metadata} features become available.

(5) Built on top of existing autoscaling algorithms, such as~\cite{cahoon2022doppler, rzadca2020autopilot}, we make adjustments to the existing customers to ``provision'' the right SKU as the reference [\textbf{C4}] to ensure accuracy and improve on satisfaction.



\subsubsection*{\textbf{Contributions}}
In sum, our contributions are as follows:
\begin{itemize}
    \item We develop an innovative prediction method for SKU recommendation based on \edit{\metadata} that leverages ``similar” customers who have already been onboarded to the cloud service.
    \item We provide two algorithms (hierarchy traversing and target encoding) to enable flexibility and explainability.
    \item We provide personalization by incorporating an online learning mechanism to constantly adjust the recommendation policy for each customer by capturing signals about customer satisfaction along the price-performance trade-off.
    \item We evaluate \sysname using production workload traces from \pg as well as synthetic workloads. Compared to our baseline, \sysname reduces wasted capacity by over 60\% without increasing throttling. Validation on synthetic signals also confirms that \sysname's personalization stage converges to align with customer preferences.
\end{itemize}

The remainder of this paper is organized as follows. 
Section \ref{sec:motivate} describes the cloud resource that \sysname is currently applied to and evaluated upon.
Section \ref{sec:lorentz} presents the overall architecture and each module in detail, with implementation details provided in Section \ref{sec:implement}.
Section \ref{sec:experiments} shows experiment results. Section \ref{sec:related} overviews related work, and, Section \ref{sec:conclusion} concludes the paper and discusses future work.

\section{Background}
\label{sec:motivate}
Although \sysname is applicable to any cloud resource with available customer \edit{\metadata}, we demonstrate it in this work for \pg (flexible server). In this section, we discuss this database system and the data it exposes to \sysname.

\subsection{DB SaaS} 
\msft is investing in a 5x5 developer experience for Database (DB) Software-as-a-Service (SaaS). One of the keys to this experience is abstracting the provisioning and tuning of DB configurations, such as the SKU, to provide knob-free services. Estimating the necessary hardware capacity for DBs is a common pain point for developers, especially for newly created workloads. The current \pg (flexible server) configuration tool merely provides the minimum-capacity SKU as the default choice, placing the burden of SKU selection entirely on users. Thus, the stated goal of \sysname is: 

\begin{quote}
\textit{"There will be no capacity input/usage prediction from end users needed during DB provisioning. The capacity management for the database instance will all be internally managed."}
\end{quote}

That is, \sysname must recommend initial virtual machine (VM) SKUs upon DB creation. 

Existing works such as \doppler~\cite{cahoon2022doppler} and Autopilot~\cite{rzadca2020autopilot} recommend DB capacities based on observed workloads: once a DB is up and running, these systems observe that DB’s resource utilization patterns to identify future optimal capacities in various dimensions. The key innovation of \sysname is making these recommendations before the DB is initialized---i.e., without any utilization data. While DB migration and autoscaling recommenders act on historical usage patterns, the \sysname engine must recommend a capacity based only on \edit{customer \metadata} and limited database-specific \edit{profile data}.
\edit{\doppler~\cite{cahoon2022doppler} focuses on workload migration where the target database's traces are available. Compared to \sysname, both methods are grounded in insights gleaned from existing provisioned databases. However, Doppler requires extensive workload telemetry from the target database as input, rendering the method impractical for our scenario.}

\sysname is generalizable to capacity recommendation for all compute resources, not just DB services. For simplicity and ease of generalization, Section~\ref{sec:lorentz} illustrates the application of \sysname on capacity recommendation for VM resources in general. This is purely a vocabulary decision---we still evaluate \sysname’s efficacy solely on the set of \pg databases. 
In a similar vein, we implement \sysname in a generalizable way, allowing the system to be configured to the specific needs of arbitrary capacity recommendation tasks.

For \pg at \msft, services are \textit{stratified} into three \textit{server offerings}, each catering to different use cases: Burstable (dev), General Purpose (small production), and Memory-Optimized (large production). The respective offerings induce distinct sets of candidate capacities:
\begin{align}
    \text{Burstable: }& \{1, 2, 4, 8, 20\} \nonumber\\
    \text{General Purpose: }& \{2, 4, 8, 16, 32, 48, 64, 96, 128\} \nonumber\\
    \text{Memory-Optimized: }& \{2, 4, 8, 16, 20, 32, 48, 64, 96, 128\} \nonumber
\end{align}
Burstable, General Purpose, and Memory-Optimized VMs make up 5\%, 49\%, and 46\% of the user database, respectively. Throughout this work, all statistics and performance metrics describe the global average across all three server offerings. Since provisioning preferences vary significantly across the three offerings, we train a distinct parameter set per offering for all recommendation models. \sysname also assumes that server offering is pre-selected, since users select their desired offering before choosing resource capacity.
\subsection{Data} 
\label{sec:data}

\sysname requires three main data categories to recommend DB SKUs: (1) Utilization telemetry for similar existing provisioned DBs, (2) \edit{\metadata} tags for existing and new DBs, and (3) signals of customer satisfaction. This section details each of the required data, and summarizes the \pg (flexible server) l0dataset we use to demonstrate \sysname. Our method is generalizable to any DB, VM, or other system that surfaces the necessary data described here.

\subsubsection*{\textbf{Telemetry for existing DBs}} 
Existing DBs in the system will be used as reference points when provisioning for new DBs. For each resource dimension (e.g., CPU, memory, IOPS), \sysname requires the following descriptive data: (1) the available capacity options (i.e., SKUs) for the service, (2) the currently selected capacity for each existing DB, and (3) telemetry tracking resource consumption for each existing DB.
In sum, \sysname requires:
\begin{itemize}
    \item \textbf{Capacity/SKU options.} We use a table containing regularly-updated SKU selections for each \pg to derive capacity recommendation options and current resource capacities.
    \item \textbf{Currently-selected SKUs.} \sysname references existing DBs' SKU choices when recommending new configurations.
    \item \textbf{Resource utilization telemetry.} For existing DBs, we require compute utilization telemetry with at a high resolution to validate (or adjust) chosen SKUs.
\end{itemize}

\subsubsection*{\textbf{\edit{Profile Data}}} 
\textit{Profile data} refers to any categorical variable describing a customer or DB instances. At \msft, the billing team records such data using identifiable tags for every chargeable cloud resource created. This can include software versions, localization tags (e.g., region or country in which the compute instance resides), or development/test/production tags. 

\edit{The profile data consists of: (1) individual database-level details, including the database ID, resource group name, and subscription ID. This information is essential for billing, ensuring that usage is attributed to a specific account; (2) Comprehensive details about each customer, such as industry and segment names. This data serves multiple purposes, aiding in marketing efforts and facilitating customer management. Establishing detailed user profiles is a common industry practice for supporting various analyses. For example, companies like Uber and Lyft~\cite{sawhney2019uber} utilize detailed customer profiles, including regional information, interviews, and behavior patterns, to inform promotion strategies and pricing regimes. Similarly, Facebook automatically tags individual users with learned preferences, such as political stances~\cite{kushin2009getting}. }

\edit{
Note that these preferences are often acquired through automated processes rather than manual logging, various methods, including machine learning, can be leveraged. Database providers like Salesforce~\cite{sf}, AWS~\cite{aws}, and Oracle~\cite{oracle} offer professional software solutions to manage such customer profile data. At \msft, this information is stored in an internal database, ensuring secure access through authentication, and facilitating downstream analysis.}

For resource provisioning on \pg, \sysname utilizes the requested DB's SKU family (i.e., server offering) and resource group (usually created to support a particular application or project), along with the hierarchy of customer \edit{\metadata}, from granular unique service account IDs (i.e., subscription ID) to broad customer segmentation tags like industry name (e.g., Food and Drink). 

In sum, \sysname leverages the following \edit{\metadata}:
\begin{itemize}
    \item \textbf{Resource Tags.} \sysname pulls resource-specific tags like software version or selected server offering from the same tables as capacity and utilization.
    \item \textbf{Customer \edit{Profile Data}.} Low-level customer \edit{\metadata} (e.g., service account or subscription ID and resource groups) are inferred from DB instance ID paths in the utilization table. Higher-level \edit{\metadata} comes from the account's metadata table from the Billing department. 
\end{itemize}

\subsubsection*{\textbf{Customer Satisfaction Signals}} 

Customer satisfaction signals comprise anything that indicates (1) a cost sensitivity—a customer prefers cheaper offerings and is willing to take slight performance hits to reduce cost—or (2) a performance sensitivity—a customer prefers more performant offerings and is willing to pay more to ensure they never experience any throttling. Examples include support tickets and manual scale actions on resources within the service.

For \pg's resource provisioning, \sysname utilizes performance and cost-related Customer Report Incidents (CRIs), which are currently labeled via a manually-crafted keyword search. We envision that more advanced CRI processing should use Large Language Models (such as a member of the OpenAI GPT family of models~\cite{dale2021gpt}) to more accurately assess the strength of perf/cost signals from each ticket.

\edit{The CRI data serve as indicators providing insights into customer preferences. Additional signals may be derived from the feedback loop, incorporating real usage and error data, as well as a sensitivity analysis comparing customer performance to cost. This analysis considers the actions of existing and new customers in adjusting capacity for this workload and similar cohorts, with these user operations being routinely recorded.}

\edit{
In our preliminary dataset comprising approximately 4,400 CRI tickets submitted by \pg users, the algorithm discerns approximately 2,400 instances of neutral sentiment (0), around 2,000 cases of performance-sensitive sentiment (1), and 5 instances of price-sensitive sentiment (-1). This aligns with our expectations, indicating that a significant portion of these CRIs are expressions of dissatisfaction related to throttling via CRI data—a more prevalent scenario compared to capacity reduction.}



\subsubsection*{\textbf{Dataset Summary}} 
\label{sec:dataset}

For each DB provisioned on \pg with a specific SKU, we notate the following information: 
\begin{itemize}
    \item $\mathbf{u}(t)$: usage data for all the resource dimensions of interest (e.g., vCores or memory), accessed from telemetry; formatted as a high-resolution time series of vectors, and each entry $u_r(t)$ corresponds to one resource dimension $r$;
    \item $\mathbf{x}$: a \edit{feature} vector describing the DB and its customer (e.g., by the customer ID, industry name, etc.);
    \item $\mathbf{c}^0$: the DB's customer-selected SKU, i.e., the SKU of the VM that hosts the service, depicted as a vector indicating the amount of resources provisioned, e.g., [4vCores, 5GB] for vCores and memory respectively.
\end{itemize}

\edit{Note that for any target (new) DBs to be provisioned, only \edit{\metadata} is available while for existing provisioned DBs, both usage data and \edit{\metadata} is available.}
We let the matrix $\mathbf{X}$ denote all the \edit{\metadata} records across DBs, where each row contains the \edit{\metadata} $\mathbf{x}$ for a particular DB. 


The dataset contains 77,584 DBs, each corresponds to one VM. And we select 7 of the \edit{\metadata} features derived in Section~\ref{sec:experiments}. Workloads $\mathbf{u}(t)$ describe at most 7 days of resource usage, sampled approximately once per minute. 

For the analyzed system, we found that:
 \begin{itemize}
     \item Users select the ideal resource capacity only 43\% of the time: 19\% over-provision and 38\% of the time under-provision, relative to \sysname's rightsized capacities (see Section 3.2).
     \item Users are approximately equally good at selecting capacities for development and production DBs.
     \item For development DBs: 
     \begin{itemize}
         \item Users under-provision for 54\% of DBs, but over-provision only 6\%.
         \item Users choose the minimum (default) capacity for 80\% of dev DBs, but it is only appropriate for 38\% of them. Note that this is likely because the default choice presented to users is the minimum capacity; across all servers, users select this choice 63\% of the time.
     \end{itemize}
     \item For production DBs:
     \begin{itemize}
         \item Users make balanced capacity guesses for production DBs 
Users are correct for 45\%, under-provision for 25\%, and over-provision for 30\% of prod DBs.
     \end{itemize}
 \end{itemize} 

\begin{figure}
    \centering
    \begin{subfigure}{0.45\textwidth}
        \hspace{-0.5cm}
        \includegraphics[width=\textwidth]{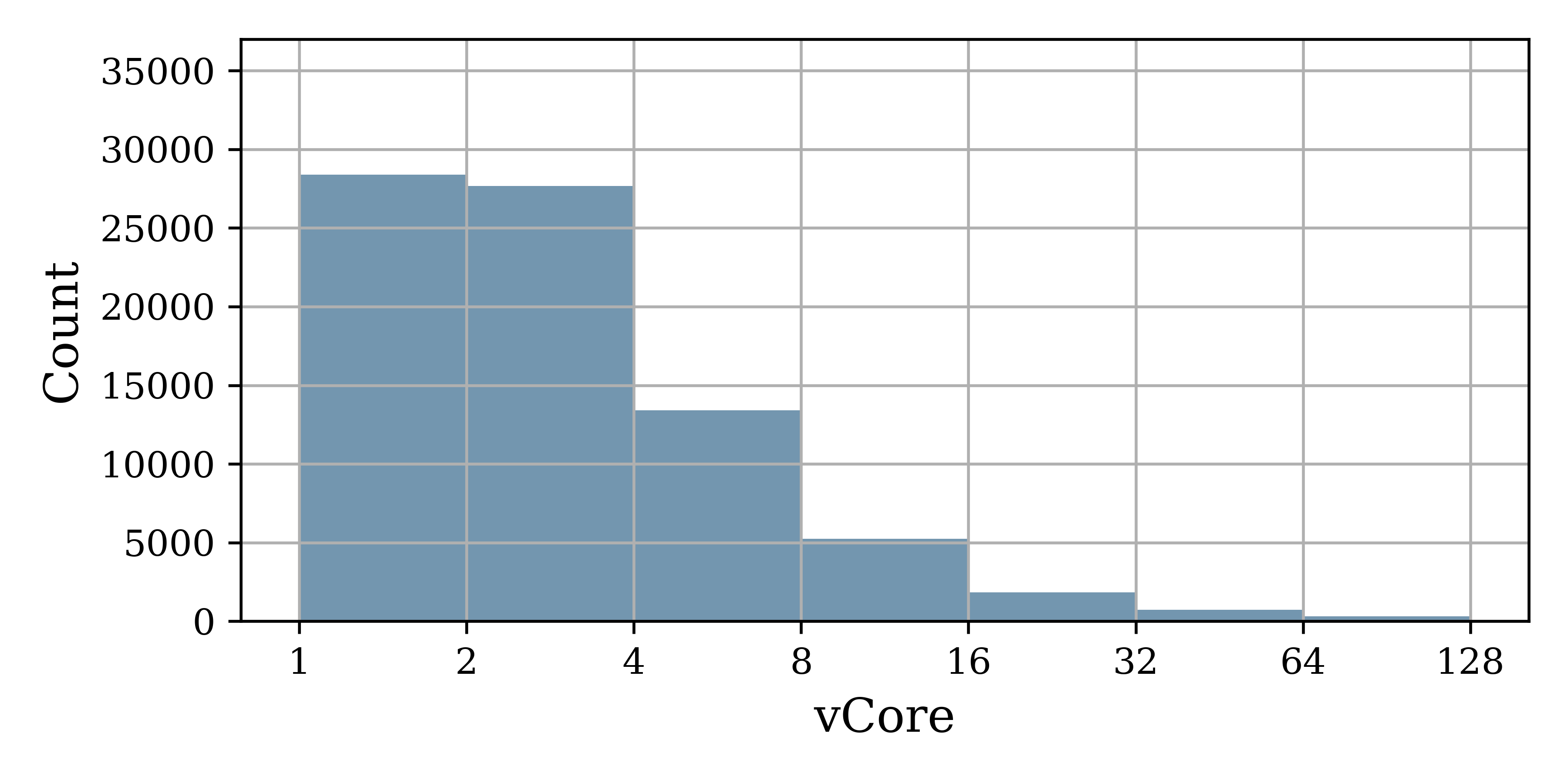}
        \caption{Distribution of user-selected vCore capacities}
    \end{subfigure}
    \vspace{0.2cm}
    \begin{subfigure}{0.45\textwidth}
        \includegraphics[width=\textwidth]{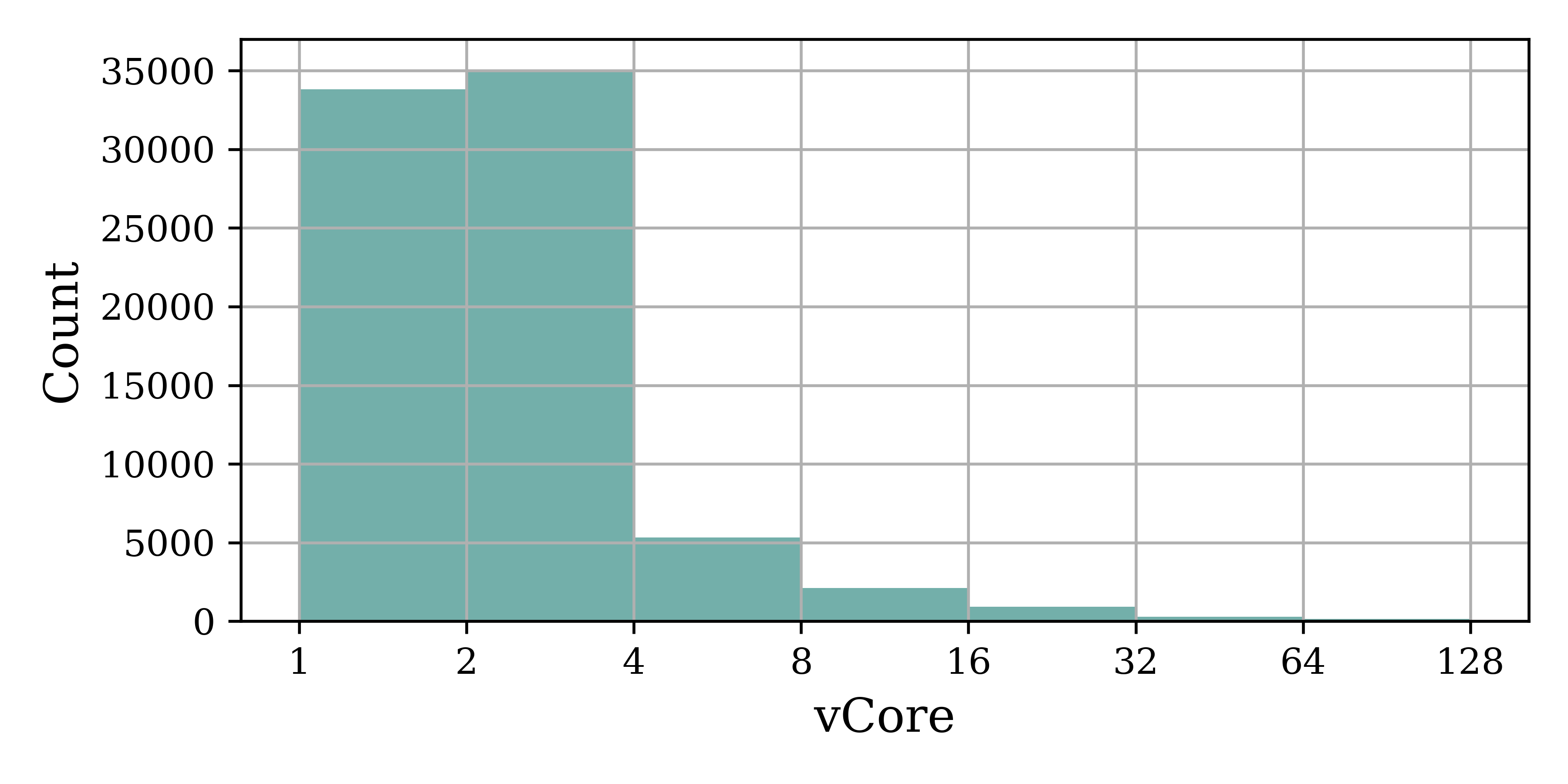}
        \caption{Distribution of rightsized vCore capacities}
    \end{subfigure}
    \caption{Rightsizing focuses the capacity distribution to prevent over- and under-provisioning.}\label{fig:skus}
\end{figure}




\section{Lorentz framework}
\label{sec:lorentz}
In this section, we describe the Lorentz framework, including the architecture of and interactions between each stage. To address the challenges listed in Section~\ref{sec:intro}, we introduce a 3-stage pipeline.

\subsection{Overview}
\begin{figure}
    \centering
    \includegraphics[width=0.6\linewidth]{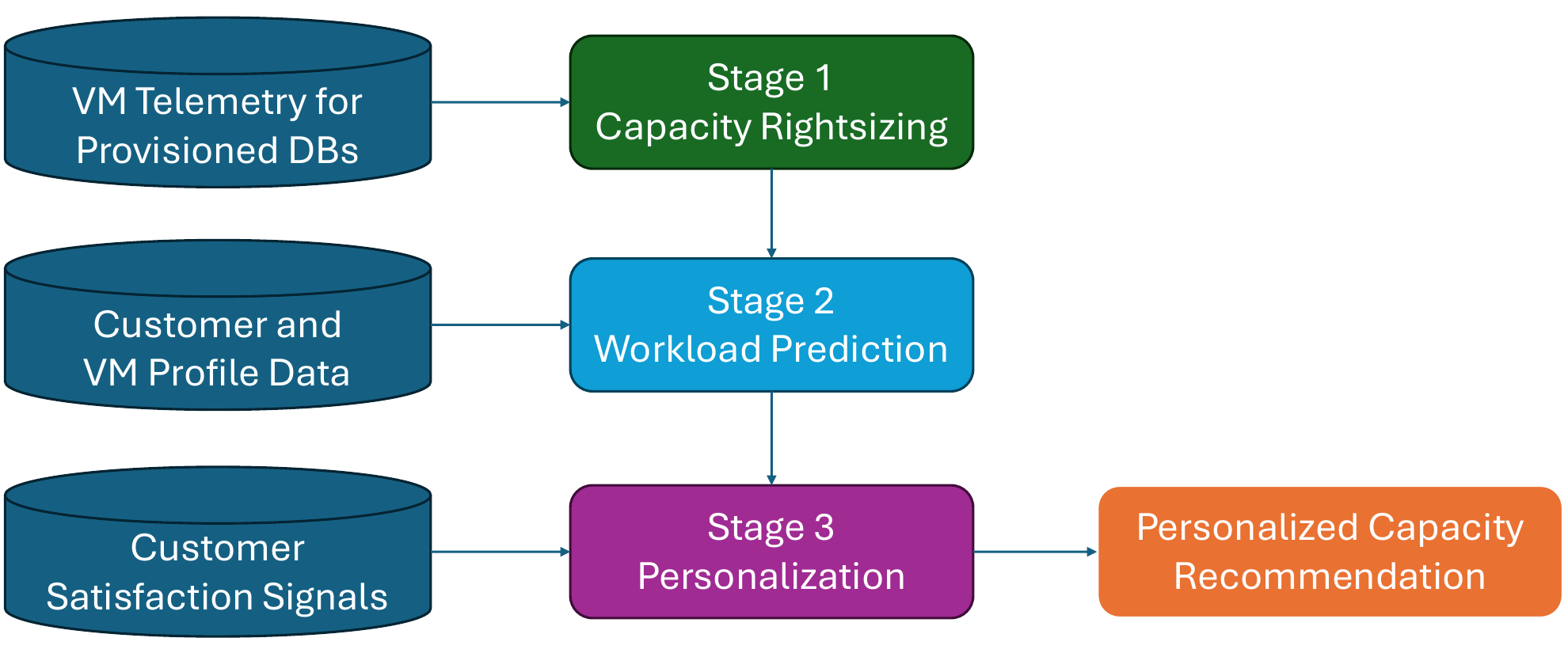}
    \caption{\edit{Inputs and outputs of each stage of the Lorentz capacity recommendation framework.}}
    \label{fig:stages}
\end{figure}

Figure \ref{fig:stages} illustrates the three stages of \sysname: Stage 1 of \sysname computes the best-fit capacity for an existing VM instance, based on its usage telemetry;
Stage 2 leverages these rightsized capacities as ``labels'', along with customer \edit{\metadata} characterized as ``features'', to recommend resource capacities for new workloads;
and, finally, Stage 3 takes arbitrary signals of customer satisfaction to continuously learn each customer's relative cost/performance sensitivity, enabling \sysname to provide personalized capacity recommendations for each customer's resource provisioning requests.

\subsection{Stage 1: Capacity rightsizing}
\begin{figure*}[t]
    \includegraphics[width=\textwidth]{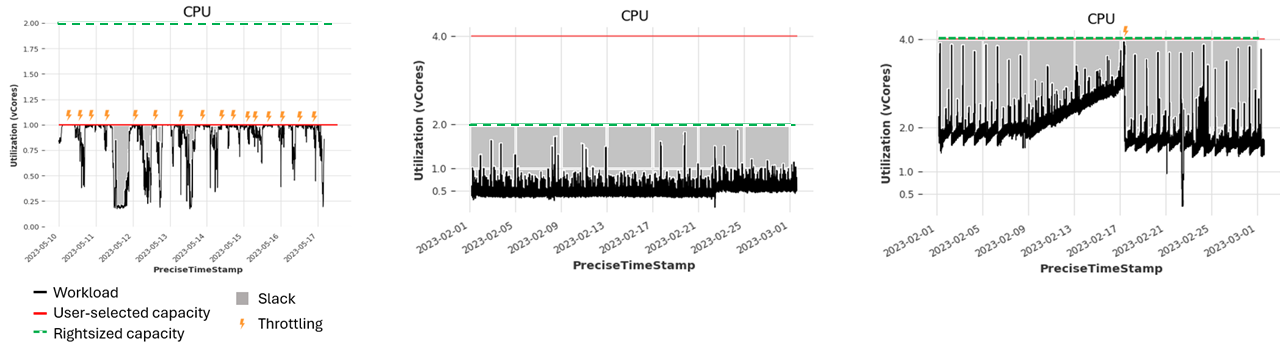}
    \vspace{-0.4cm}
    \caption{CPU slack and throttling for under-provisioned (left), over-provisioned (center), and well-provisioned (right) \pg database VMs. \edit{The dashed line shows the rightsized SKU.}}
    \label{fig:rightsizing}
        \vspace{-0.2cm}
\end{figure*}

Stage 1 begins with rightsizing the SKUs for existing users, especially for the ones that are over- or under-provisioned \edit(see Figure~\ref{fig:rightsizing}). 
In the telemetry data collected for existing workloads, let $\textbf{u}(t)$ denote the irregularly sampled time series that measures the raw resource utilization for a DB hosted on a VM (up to the user's selected capacity, denoted as $\textbf{c}^0$). We have thus:
\begin{align}
    u_r(t) \leq c_{r}^0, ~\quad \forall t, r,
\end{align}
where $t$ is the time stamp and $r$ the resource dimension index, such as CPU, memory, and IOPS.

First, to standardize its temporal resolution, we group the irregularly-spaced samples $\textbf{u}(t)$ into $T$-minute bins and aggregate each bin to a single observation, producing a regular workload resource usage signal $\textbf{w}[n]$ with fixed time intervals. Our use case usually samples utilization approximately every minute, and we select $\text{max}(\cdot)$ as our aggregator to measure worst-case performance \edit{thus avoiding under-provisioning.
\begin{align}
    \textbf{w}[n] &= \text{max}(\{\textbf{u}(t) | n \leq \frac{t}{T} < n+1\}).
\end{align}
}
To compute the rightsized capacity of existing DBs using the regularly sampled workload trace $\textbf{w}[n]$, we design a selection criterion that selects the optimal SKU candidate from a set of candidate capacities $\mathcal{C}$, based solely on the workload resource usage. We first compute two opposing statistics describing the resource utilization: \textit{throttling} and \textit{slack} for each SKU candidate $\textbf{c}$, and then define an optimizer that selects the capacity that best balances the two statistics for the rightsizing.

\subsubsection*{\textbf{Throttling}}
When the selected capacity is too small to accommodate the workload, \textit{throttling} occurs; we call this resource \textit{under-provisioned} (see left of Figure~\ref{fig:rightsizing}\edit{)}. For some resources (e.g., memory), this can result in the cancellation of jobs (e.g., out-of-memory errors); this is nearly always an undesirable behavior. For other resources like CPU, jobs are simply delayed until the requisite resource space is available, which can more often be accommodated. Tolerance for throttling may vary across use cases and resource dimensions; for example, production services with tight reliability and latency requirements may be particularly sensitive to throttling, while minor throttling for CPU can be more tolerable than memory.

To compute throttling for SKU candidate $\textbf{c}$ given the workload with resource usage $\textbf{w}[n]$, we first set a ratio threshold $\eta_r$ of resource utilization for each resource dimension, above which performance degradation can be expected. This value should be less than 1. This threshold also indicates the optimal utilization level that users hope to achieve while minimizing the amount of unused resources without impacting performance.
Given resource usage of a workload, $\textbf{w}[n]$, we define the probability of throttling $T_w(\textbf{c})$ under candidate capacity $\textbf{c}$ as:
\begin{align}\label{eq:throttling}
    t_w[n] &= 
        \begin{cases}
            1 & \text{if } \exists r \text{ s.t.,} ~ w_r[n] > \eta_r c_r \\
            0 & \text{otherwise} \\
        \end{cases}, \\
    T_w(\textbf{c}) &= \frac{1}{n}\sum\limits_n t[n],
\end{align}
where $T_w(c)$ captures the percentage of time frames where the resource is under-provisioned (there exists one resource dimension $r$ whose utilization is above $\eta_r$). 


\subsubsection*{\textbf{Slack}}
Whenever a resource capacity is larger than necessary, some of the available resources go unused; we refer to this unused capacity as \textit{slack} and call the resource \textit{over-provisioned} (Figure \ref{fig:rightsizing}, middle). In practice, the presence of some slack is unavoidable, since it is not feasible or efficient to scale resource capacity to exactly match the needs of upcoming jobs, even if those needs could be perfectly predicted. Still, users aim to minimize slack, since an excess of this wasted space incurs costs that would have been avoidable with better capacity selection. 

We define the average slack ratio vector $\textbf{S}_w(\mathbf{c})$ induced by picking candidate capacity $\textbf{c}$ on workload $\textbf{w}[n]$ as follows, where each entry corresponds to one resource dimension $r$: 
\begin{align}
   s_{w,r}[n] &= \frac{c_r - w_r[n]}{c_r}, \\
   S_{w,r}(\mathbf{c}) &= \frac{1}{n}\sum\limits_ns_{w,r}[n],\label{eq:slack}
\end{align}
where $\textbf{S}_w(\mathbf{c})$ captures the percentage of resources that will be unused given the provisioned resource amount of $\textbf{c}$, averaged over time, and can be estimated for each resource dimension respectively.

Note that the observed resource usage is always bounded by the original user-selected capacity $\textbf{c}^0$ from the telemetry data, and $0 \leq w_r[n] \leq c^0_r~\forall r,n$. 
Therefore, in instances of throttling, the true requested quantity of the resource is censored by the capacity chosen by the user. 
In such cases where $\textbf{c} > \textbf{c}^0$, $\textbf{S}_w(\mathbf{c})$ as computed in Equation~\eqref{eq:slack} is thus an overestimation of the true slack.
To account for this situation, slack computed using Equation~\eqref{eq:slack} is no longer the critical selection criterion in the process of rightsizing when the observed workload has been throttled (see Equation~\eqref{eq:throttledrightsizing}). 

\subsubsection*{\textbf{Basic rightsizing optimizer}}

Slack is directly proportional to resource capacity, while throttling is inversely related to capacity; it is rarely possible to achieve both near-zero slack and zero throttling. We cast the navigation of this slack/throttling trade-off as an optimization problem.

Our optimizer must select the best capacity choice from the SKU candidate set for workload $\textbf{w}$: $\hat{\textbf{c}}^0(\textbf{w}) \in \mathcal{C}$. Users generally have stricter requirements for throttling (e.g., four-nines reliability) than for slack, so we construct our optimization function as a slack target $\textbf{s}^*$ subject to a throttling constraint (upper bound) of $\tau$. Especially for un-censored data (where the workload was not throttled), the rightsized capacity $\hat{c}_r$ for \edit{each resource dimension $r$ is as follows:
\begin{align}
    \hat{c}_r^0(\textbf{w}) = \underset{\textbf{c} \in \mathcal{C}}{\text{argmin}} \lvert S_{w,r}(\mathbf{c}) - s^*_r \rvert , \quad \text{s.t.}, T_w(\textbf{c}) \leq \tau, \forall r, \label{eq:rightsizing}
\end{align}
} where $s^*_r$ represents the optimal slack we hope to preserve for each resource dimension and the maximum percentage of throttling $\tau=0$ in this work as a conservative, throttling-averse starting point that can be calibrated through personalization. \edit{Combined with Equation~\eqref{eq:throttling}, we provide two alternatives for resource reprioritization. In Equation~\eqref{eq:throttling} where we define throttling as a ratio $\eta_r$ of total capacity for each resource  dimension $r$ independently, if throttling e.g. disk, is undesirable, its $\mu_r$ can be lower. Second, the slack target $s^*_r$ in Equation~\eqref{eq:rightsizing} can be tuned, enforcing low targets for costly resources like memory and higher targets for cheaper ones like disk space.}


\subsubsection*{\textbf{Rightsizing throttled workloads}}
An additional effect of the workload censoring described previously is that both the true slack and the throttling probability for candidate capacities $\textbf{c}$ larger than the user-provisioned capacity $\textbf{c}^0$ cannot be exactly computed, due to the censored observed resource usage level. Resource requests larger than $\textbf{c}^0$ are censored to equal $\textbf{c}^0$, and $T_w(\textbf{c})$ will always be $\textbf{0}$ when $\textbf{c} \geq \textbf{c}^0$, and the slack is overestimated. 

\edit{When throttled, we assume that the rightsized capacity should be at least $2^K$ times the current capacity. The calibration of K can be done by examining the telemetry data of formerly throttled servers whose capacities have been scaled up.}
To rightsize censored workloads, we use the following optimizer instead, $\forall r$:
\begin{align}
    \hat{c}_r^0(\textbf{w}) = \underset{\textbf{c} \in \mathcal{C}}{\text{argmin}}  \lvert S_{w,r}(\mathbf{c}) - s^*_r \rvert,  \quad \text{s.t.}, c_r \geq 2^K c_r^0.\label{eq:throttledrightsizing}
\end{align}

\subsubsection*{\textbf{Complete rightsizing optimizer}}
Combining the censored and uncensored cases, we define the following complete optimizer, the rightsized SKU $\hat{\mathbf{c}}^0(\textbf{w})$ is computed as:
\begin{align}
    \hat{c}^0_r(\textbf{w}) = \begin{cases}
            \underset{\textbf{c} \in \mathcal{C}}{\text{argmin}}  \lvert S_{w,r}(\mathbf{c}) - s^*_r \rvert , \quad \text{s.t.} , T_w(\mathbf{c}) \leq \tau & \text{if } T_w(\mathbf{c}^0) = 0,\\
            \underset{\textbf{c} \in \mathcal{C}}{\text{argmin}}  \lvert S_{w,r}(\mathbf{c}) - s^*_r \rvert , \quad \text{s.t.} , c_r \geq 2^K c^0_r  & \text{otherwise}. \\
        \end{cases} 
\end{align}


As we describe in the following section, the rightsized capacities that this optimizer computes become useful labels for training models that recommend resource capacities for new workloads. For \pg, we focus on the analysis for the \textit{virtual cores} (vCores, i.e., CPU) resource as the capacity for memory is provisioned proportional to the number of vCores, e.g., 4GB per core. And the estimation can be simplified as CPU constraints mostly dominate.

\subsection{Stage 2: Capacity recommenders}
\sysname's second stage recommends resource capacities for a newly-requested VM for the database, based on resource capacities of VMs performing similar duties. We provide two alternative models for this task, which we refer to as \textit{provisioners}. Since the VMs that \sysname provisions for are new, they have no telemetry data for workload traces upon which to compute VM/workload similarity; instead, we use customer- and server-level \edit{\metadata} as input features. We leverage existing users' provisioned SKUs and their \edit{\metadata} as the training dataset. For labels of this training dataset, we use the rightsized capacities computed in Stage 1 from provisioned resources $\hat{\mathbf{c}}^0$ (as opposed to user-selected SKUs $\mathbf{c}^0$), since they are conceptually and empirically an improvement over user-selected capacities. 

\subsubsection*{\textbf{Notation}}

Let $\textbf{x} \in \mathbb{R}^{M_c + M_s}$ be a vector with $M_c$ customer-level and $M_s$ server-level \edit{\metadata} features for a newly-requested VM. The goal of provisioner models is to provide a function mapping $f: \mathbb{R}^{M_c + M_s} \rightarrow \textbf{c} \in \mathcal{C}$ for the recommendation. Let $\mathbf{c}^*$ denote the recommended SKU candidate for Stage 2, we have: $\mathbf{c}^* = f(\textbf{x})$. 

\paragraph{Transformations} For regression methods, the exponential scale of many compute resources (e.g., vCores $\in\{1,2,4,8,16,...\}$) can lead to undesirable statistical properties, including heteroskedastic noise~\cite{noise}, which in turn hurts prediction efficiency. We thus fit our models in a transformed space, e.g., $\tilde{c_r} =\log_2(c_r)$, then invert the transform for our predictions.
We generally denote this transform and its inverse as $\xi(\cdot)$ and $\xi^{-1}(\cdot)$.

\subsubsection*{\textbf{Hierarchical provisioner}}

Our first provisioner model is a heuristic method that explicitly leverages hierarchical relationships in the input \edit{feature} to make recommendations. For instance, in the observed \edit{\metadata}, one customer account (referred to as CloudCustomerGuid) can have multiple subscription IDs, which can further have multiple resource groups, etc. We want to learn the hierarchy of all the entries (including both customer- and server-level features) in the \edit{\metadata} automatically. Thus we train the hierarchical provisioner with two steps:
\begin{enumerate}
    \item Compute the hierarchical structure of the service's observed \edit{\metadata} features; 
    \item Sort each observed resource capacity into "buckets" along the computed hierarchy's feature values, similar to building a decision tree for the inference.
\end{enumerate}
Inference is then performed by selecting the bucket best fitting the new cloud resource's \edit{\metadata}, and recommending a capacity based on samples resigned in that bucket.

To compute the \edit{\metadata} hierarchy, we first leverage the HALO entropy-based approach~\cite{zhang2021halo} to identify the strength of hierarchical relationships between each pair of \edit{profile} features. \edit{A Python implementation of the paper can be found in~\cite{halogithub}.} Although the HALO approach was designed with strict hierarchies (where each node except the root has exactly one parent node) in mind, we find it effective for our non-strict case as well.

\begin{figure}[t]
    \centering
    \includegraphics[width=0.6\linewidth]{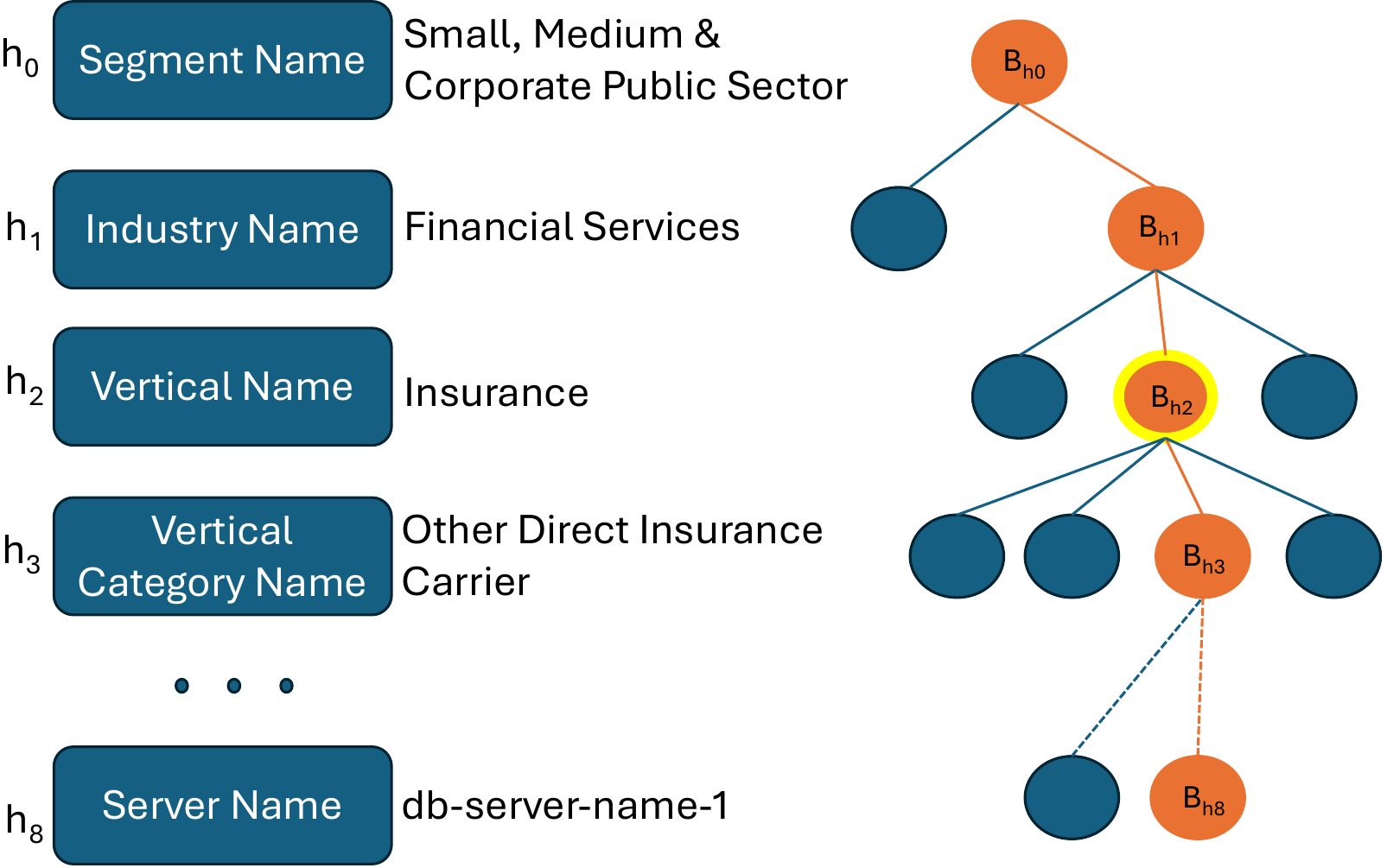}
    \caption{\edit{Example of the learned hierarchy and buckets.}}
    \label{fig:bucket}
\end{figure}

\eat{
To establish the pairwise hierarchy strength estimation between features $m_1$ and $m_2$ ($0\leq m_1,m_2 < M_c + M_s$), we first compute the conditional entropy between them~\cite{zhang2021halo}:
\begin{equation}
    H(m_1|m_2) = -\sum\limits_{u \in \mathbf{X}_{\cdot,m_1}, v \in \mathbf{X}_{\cdot,m_2}} p(u, v)\text{log}\left(\frac{p(u, v)}{p(v)}\right)
\end{equation}

Next, we compute the uncertainty reduction $UR$ for $m_1$ given $m_2$ as:
\begin{equation}
    \text{UR}(m_1|m_2) = 1 - \frac{H(m_1|m_2)}{H(m_1)}
\end{equation}

where $H(m_1)$ is the entropy of feature $m_1$: 
\begin{equation}
    H(m_1) = \sum\limits_{u \in \mathbf{X}_{\cdot,m}}p(u)\textnormal{log}\left(p(u)\right)
\end{equation}

Finally, we compute a matrix of hierarchy intensity values \text{HI} $\in \mathbb{R}^{M_c + M_s \times M_c + M_s}$ between each pair of \edit{profile} features~\cite{zhang2021halo}. We use a threshold $\gamma$ to ensure that only meaningful relationships are included in the final hierarchy. 
This matrix's entries are defined as follows:

\begin{align}
    \text{HI}_{m_1, m_2} = \begin{cases}
            \text{UR}(m_2|m_1) & \text{if } H(m_2) < H(m_1) \land UR(m_2|m_1) \geq \gamma \\
            0 & \text{otherwise}
        \end{cases} 
\end{align}

Note that $0 \leq \text{HI}_{m_1, m_2} \leq 1$, where 0 represents no hierarchical relationship and 1 represents a strict hierarchical relationship (i.e., the value of more granular feature $m_2$ exactly determines the value of coarse feature $m_2$). Each feature pair can only have a one-way relationship; that is, at most one of $\text{HI}_{m_1,m_2}$ and $\text{HI}_{m_2,m_1}$ can be non-zero~\cite{zhang2021halo}.

    Finally, }
To compute the hierarchy chain $\mathbf{h}$, we construct a weighted directed acyclic graph (DAG), using the thresholded table as the adjacency matrix; each edge points toward the coarser node. We then select the node with the highest out-degree as the root $\mathbf{h}_0$ of the hierarchy and traverse the graph by selecting the current node's highest out-degree neighbor until a node with out-degree 0 is reached~\cite{zhang2021halo}.
\edit{In the DAG, lower granularity features always point towards higher granularity features.}
Figure~\ref{fig:bucket} shows an example of the learned hierarchy, i.e., \edit{SegmentName > IndustryName > VerticalName > VerticalCategoryName > ... > ServerName,
where for samples share the same value for the child node features (e.g., both have $h_2$=`Insurance'), they also share the same value for the corresponding parent node (e.g., both have $h_1$=`Financial Services'. }

Based on the learned hierarchy of \edit{profile} features, the next step is to populate a set of buckets $B$ along the computed hierarchy. For each feature $\mathbf{h}_j, 0 \leq j < |\mathbf{h}|$, we define one level of buckets $B_{\mathbf{h}_j}$ (see Figure~\ref{fig:bucket}); within each level, we initialize one bucket $B_{\mathbf{h_j}, v}$ for each unique value $v$ of feature $\mathbf{h}_j$. We populate each bucket with the capacities of existing provisioned VMs based on the \edit{\metadata} for each with that bucket's feature value\footnote{Due to user mis-entry of \edit{profile} features, the hierarchies in our dataset are nearly, but not perfectly strict (i.e., HI values slightly less than 1). As a result, it is most appropriate to index bucket $B_{\mathbf{h}_j, v}$ by a single hierarchy level $\mathbf{h}_j$ (instead of including all features coarser than $\mathbf{h}_j$), since this eliminates noise in the coarser features.} such that:
\begin{align}
    & B_{\mathbf{h}_j, v} = \{\hat{\mathbf{c}}^0_i | x_{i, \mathbf{h}_j} = v \}.
\end{align}

To recommend the capacity for a newly-provisioned VM resource with \edit{profile feature} vector $\mathbf{x}$, we first select its most granular feature $\mathbf{h}_k$ whose feature-value pair has a sufficient bucket size at the corresponding level, i.e., $\geq N$. We then return the $p^{\text{th}}$ percentile of the observed capacities in that bucket as the recommendation $\mathbf{c}^*$:
\edit{
\begin{align}
    k &= \underset{l}{\text{argmax}}(|B_{\mathbf{h}_l, \mathbf{x}_l} |\geq N), \\
    \mathbf{c}^* &= f(\textbf{x}) = \text{\%ile}(B_{\mathbf{h}_k, \mathbf{x}_k}, p), \label{eq:bucket}
\end{align}}
where $\mathbf{c}^*$ is the recommended SKU candidate from Stage 2 using the mapping function $f$.



Figure~\ref{fig:bucket} provides an example of inference. The \edit{\metadata} of a hypothetical VM indicates a particular sequence of buckets it belongs to, following the computed hierarchy \textbf{h}. \edit{Starting from \textbf{h}$_8$ (ServerName), we might find that bucket $B_{\textbf{h}_8, \text{db-server-name-1}}$} is not sufficiently large to make a recommendation. In this case, we traverse up the hierarchy until reaching $B_{\textbf{h}_2, \text{Insurance}}$, the first sufficiently-large bucket. We observe from this bucket the distribution of rightsized capacities $\hat{\textbf{c}}^0$ (computed in Stage 1) for all servers in the ``Insurance'' vertical, taking the $p^{\text{th}}$ percentile as the capacity recommendation.

\subsubsection*{\textbf{Target encoding provisioner}}
Our second provisioner model uses target encoding (TE) to admit non-parametric regression or classification ML models like tree-based models \edit{which are best on tabular data~\cite{grinsztajn2022tree}}, such as LightGBM ~\cite{ke2017lightgbm}. This approach is suited to (i) data with non-hierarchical relationships between \edit{\metadata} tags and (ii) large enough data size that arbitrary relationships between variables can be learned by non-parametric models. While the framework technically admits arbitrary categorical variable encodings and regression/classification methods, we choose tree-based predictors and TE because (i) tree-based predictors offer best-in-class performance on tabular data and (ii) alternative encodings--such as one-hot encoding--often behave poorly with trees and high-cardinality \edit{\metadata} tags.

Target encoding converts every categorical feature into a numeric using statistics from the whole training data. For instance, in the column/\edit{\metadata} entry of SegmentName, one particular sample has the value ``Beverage''. This value is replaced with the average vCores (i.e., the label) from the subset of samples whose SegmentName equals ``Beverage''. The average vCore values can be later used as ML prediction features to predict the same label.
This approach can be broken into two steps:
\begin{enumerate}
    \item For each row (i.e., one VM) in the training data $X$, for each attribute at entry $h$, $x_h$, based on the SKU label $\hat{\mathbf{c}}^0$ (after rightsizing) of this VM and feature value $x_h=v$ such as ``Beverage'', we map $\text{TE}(x_h)=\psi_{\{n|X_{n,h}=v\}}(\textbf{c}_n)$, where $\psi(\cdot)$ is some aggregation function, such as a mean or percentile. The set $\{n|X_{n,h}=v\}$ includes all the observations of VMs whose \edit{profile} feature at entry $h$ equals the observed value $v$. $\text{TE}(x_i)$ estimates the statistics of the labels from the subset of samples given a particular attribute value.
    \item After transforming all the features in all observations $\textbf{x} \in X$ (i.e., the ones we learned in the hierarchy model) with the mapping $\text{TE}(x_h)$, use a traditional regression method to estimate $\textbf{c}^*= f( \mathbf{x})$.
\end{enumerate}



\subsubsection*{\textbf{Missing data}}
The handling of missing \edit{\metadata} tags with a TE approach is non-trivial. We explored the options of replacing the tag ``missing'' in each attribute with an invalid numerical value, e.g., $-999$, but found both random forest~\cite{rf} and gradient-boosted trees~\cite{gb} unable to differentiate this case from the other encoded values, resulting in severe underestimation in the presence of missing data. This problem vanishes when replacing the tag ``missing'' with the average of $\hat{\textbf{c}}^0$ over the whole training set. 

\edit{
\subsubsection*{\textbf{Extension to include trace data}}
The method can be extended to incorporate additional data types. For instance, if trace data becomes accessible, both the hierarchical and target encoding models can utilize more features as inputs, thereby enhancing prediction accuracy. Specifically, the inclusion of more features contributes to improving the variability of SKU choices within each bucket, leading to a more powerful predictor. In this context, \sysname can serve as a predictive tool to assist in decision-making for autoscaling after resources have been provisioned.
}

\subsection{Stage 3: Personalization}\label{sec:personalization}

By training Stage 2 of Lorentz on the rightsized capacities from Stage 1, the system is able to predict a capacity fitted to the unseen workload of a new resource based on \edit{profile} features $\mathbf{x}$. However, no matter how reasonable the definition of “rightsized” may be to a given user, different users have different needs; in fact, even a single customer may have different needs across their various resource groups within a service.

Consider two types of databases for a beverage company: trucking logs and a small marketing team’s database. Faults in trucking logs may result in delayed shipments, meaning the company is willing to pay significantly more to ensure their trucking logs never experience any faults. Meanwhile, the small marketing team may never see heavy database traffic, and thus may prioritize a good deal with cheaper price.

Stage 3 of \sysname treats the rightsizing recommendations from the Stage 1 algorithm as a \textit{baseline recommendation}; that is, a typical customer may be fine with that fit of capacity to workload. We then use customer \textit{cost/performance sensitivity scores}, denoted by $\lambda$, to further adjust the output of Stages 2. A performance-sensitive customer will receive a higher capacity than the Stage 2 recommendation, while a cost-sensitive customer will receive a smaller recommendation.

\begin{figure}[t]
    \centering
    \includegraphics[width=\linewidth]{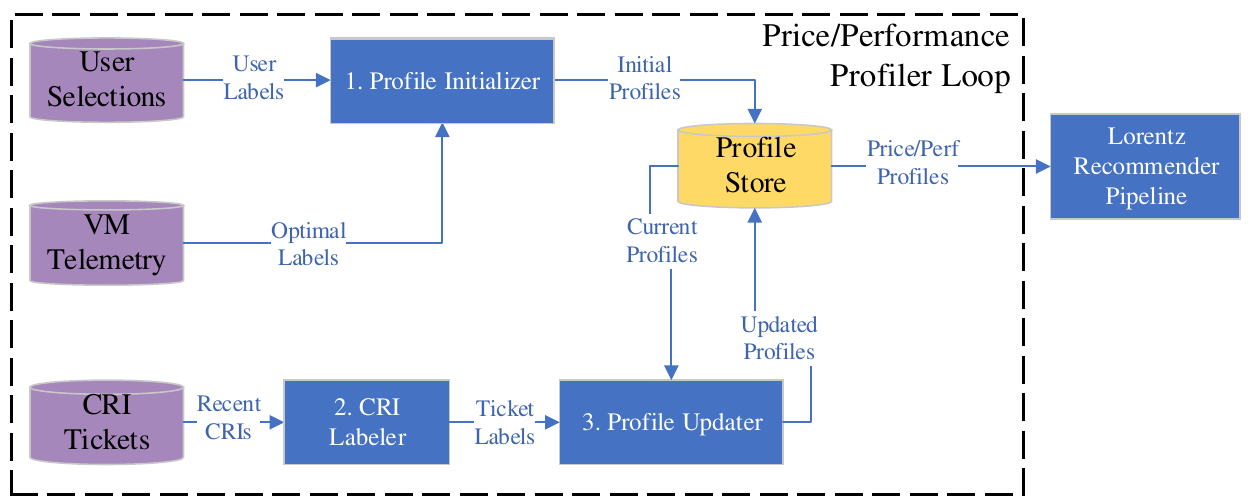}
    \caption{The personalization loop completes \sysname.}
    \label{fig:stage3}
\end{figure}

We develop the \sysname's personalization algorithm to keep a profile of each individual user (or even finer granularity) about their sensitivity to price or performance. At initialization time, such profiles can be created based on their current selection of SKUs and the resource utilization level (see Figure~\ref{fig:stage3}). A profile store is thus created and maintained to store detailed information. Whenever a learning signal is triggered, such as a customer-reported incident (CRI) is issued, we will (1) determine if such a signal is related to the performance or cost of a customer, and (2) update the customer profile accordingly. Based on the constantly learned customer profile, the output recommendation from Stage 2 will be further adjusted to accommodate the different preferences of price versus performance accordingly. As the signal can be very sparse, we develop an innovative profile update algorithm similar to message propagation in social networks~\cite{petrovic2011rt} to amplify the impacts of each received signal.

\subsubsection{$\lambda$-adjusted Recommendations}
Stage 3 uses the recommendations from Stage 2 along with a learned cost-performance sensitivity score of $\lambda$ to compute adjusted scores in a two-stage approach. We remain aware of the log-transformation $\xi(\cdot)$ in this process. We get the adjusted capacity $\mathbf{c}^{**}=g(\mathbf{c}^*;\lambda)$ from the Stage 2's recommended capacity $\mathbf{c}^*$ as:
\begin{equation}
    \mathbf{c}^{**}=g(\mathbf{c}^*;\lambda)=\xi^{-1}\bigg(\xi\big(\mathbf{c}^*\big)+\lambda\bigg).
\end{equation}
In the case of $\xi=\log_b$, $\lambda$ is easily interpreted as ``How many powers of $b$ should we increase or decrease the Stage 2's predictions by?'':

\begin{align}
\mathbf{c}^{**}=g(\mathbf{c}^*;\lambda)=b^{\left(\log_b(\mathbf{c}^*)+\lambda\right)}=b^\lambda \mathbf{c}^*.  
\end{align}

The larger $\lambda$ is the more costly offers will be recommended indicating that the user prefer a more performant service.

\subsubsection{Learning customized $\lambda$}
If a customer generates cost-sensitive signals (e.g., submits tickets to complain about cost or manually scale down resource capacity for their VMs), their cost-performance sensitivity score $\lambda$ will decrease, adjusting future recommendations down. If they complain about performance and the mitigation is to provision a larger SKU, their $\lambda$ will increase.

\paragraph{Structuring $\lambda$} 
We consider the hierarchy that can be learned from Stage 1 
as the structure to compute the $\lambda$ values, such as CloudCustomerGUID 
> SubscriptionID > ResourceGroup > Stratification for individual service or product. 
We can record the cost-perf sensitivity score $\lambda$ at any of these granularities. 
Because predictions are mostly made at the Resource Group (RG) level in Stage 2, and because RGs are typically grouped with similar purposes (and thus often similar workload patterns), we will structure $\lambda$ as a vector $\mathbf{\lambda}_{\text{C},\text{S},\text{R}}$ whose dimension equals the number of possible stratification values and who is indexed by CustomerGUID $\text{C}$, subscription ID $\text{S}$, and RG $\text{R}$. Note that, each RG has its own vector of preferences across stratification variables; thus, if there are 3 stratification variables, there will be 3 sets of $\lambda$ values in each RG. \edit{For new user profiles, $\lambda$ defaults to 0.}

\paragraph{Customer Satisfaction Signals} Stage 3 operates by adapting predictions to feedback on whether a given customer needs higher performance or lower costs. In order to capture this information, we broadly define customer satisfaction signals as any signal indicating a preference for more performant or lower-cost resources. These can include Customer Report Incidents (CRIs) with performance or cost as primary subjects, manual scale actions, or any other observable signal that can reliably indicate such a preference. We map these signals to the range $\gamma\in[-1,1]$, where $-1$ indicates a strong cost sensitivity (prefer cheaper solutions), $+1$ indicates a strong performance sensitivity (prefer performant solutions), and anything in between indicates a lesser lean in that direction.  \edit{We integrated a simple strategy for sentiment detection by checking if keywords such as ``scale-up'', ``increase CPU'', etc., exist in the ticket. Specifically, we focus on three ticket fields—``symptom", ``title", and ``resolution"—to formulate distinct sets of keywords for discerning the intent and resolution of the ticket. An example of such keywords for identifying complaints related to throttling can be found in Table~\ref{tab:keyword}. In the future, a more comprehensive analysis could be conducted, leveraging tools such as Large Language Models (LLM) or machine learning classifiers~\cite{crxclassification}.}
\begin{table}[h]
  \centering
  \caption{\edit{Throttle Filters}} \label{tab:keyword}
  \vspace{-0.2cm}
\resizebox{10.7cm}{1.5cm}{%
  \begin{tabular}{lll}
    \toprule
    Column & Keywords \\
    \midrule
    Symptoms & [`high cpu', `high cpu usage', `high cpu utilization', `high cpu utilisation'] \\
    Subject & [`high cpu', `high cpu usage', `high cpu utilization', `high cpu utilisation'] \\ 
    Subject & [`100\%', `99\%', `95\%', `90\%', `0\%', `9\%'] \\ 
    Resolution & [`increas'] \\ 
    Resolution & [`throttl'] \\ 
    Resolution & [`scale up', `scaling up', `scaled up'] \\ 
    \bottomrule
  \end{tabular}}
    \vspace{-0.4cm}
\end{table}


\paragraph{Algorithm} 
As the satisfaction signals can be very sparse, in this paper, we develop a message propagation algorithm to capture the network impact and dynamically update the related profile across services/resource groups. Each time \sysname receives a customer satisfaction signal, it propagates it into the associated customer profiles via weighted addition. For example, if a customer files a complaint about performance in a given resource group, the signal will still have a (reduced) impact on other resource groups across that customer’s subscriptions, but \textit{will not impact scores for other customers}. \edit{Algorithm~\ref{alg:message-propagation} details the process. The signal is first multiplied by a learning rate $l_r$, which controls the overall speed with which Lorentz reacts to signals. Relative impact in other buckets is controlled by multiplicative decay parameters $\rho_R$ (decay across stratifications for different server offerings, e.g., Bustable, General Purpose), $\rho_S$ (decay across Resource Groups) and $\rho_C$ (decay across Subscriptions).}
\edit{
\begin{algorithm}
\caption{\edit{Message Propagation Algorithm for Customer Sensitivity Score Update}}
\label{alg:message-propagation}
\KwIn{\edit{Satisfaction signal $\gamma$ for customer $\text{CU}$, subscription $\text{SU}$, resource group $\text{RG}$, stratification $\text{ST}$, learning rate $l_r$, decay parameters $\rho_R$, $\rho_S$, $\rho_C$, initial sensitivity scores $\lambda$}}
\For{\edit{each signal $\gamma$}}{\edit{
    Multiply signal by learning rate: $s = l_r \times \gamma$;\\
    Step 1: Update score for the same resource group and stratification: $\lambda_{\text{RG}, \text{ST}} \gets \lambda_{\text{RG}, \text{ST}} + s$;\\
    Step 2: Decay signal and add to other stratification dimensions in the same resource group: $\delta = \rho_R \times s$, $\lambda_{\text{RG}, x} \gets \lambda_{\text{RG}, x} + \delta, \forall x \in \text{RG} \setminus \text{ST} $;\\
    Step 3: Accumulate updates in other resource groups in the same subscription: 
    $\lambda_{x, y} \gets \lambda_{x, y} + \rho_S \times \delta,~ \forall x \in \text{SU} \setminus \text{RG}, y \ne \text{ST} $; or
    $\lambda_{x, \text{ST}} \gets \lambda_{x, \text{ST}} + \rho_S \times s,~ \forall x \in \text{SU} \setminus \text{RG}$;\\
    Step 4: Add decayed updates to other subscriptions under the same customer: $~~\lambda_{x, y} \gets \lambda_{x, y} + \rho_C \times \delta,~ \forall x \ne \text{SU}, y \ne \text{ST} $; or
    $\lambda_{x, \text{ST}} \gets \lambda_{x, \text{ST}} + \rho_C \times s,~ \forall x \ne \text{SU}$;\\
    }}
\end{algorithm}
}

\begin{figure}
    \centering
    \includegraphics[width=.6\linewidth]{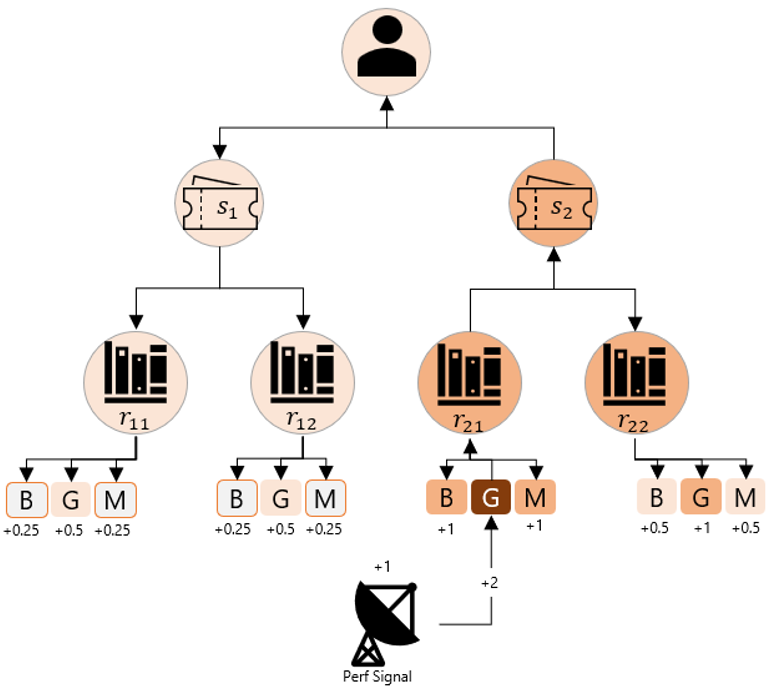}
    \vspace{-0.2cm}
    \caption{Example of updating $\lambda$ with exaggerated numbers: signal $\gamma=1$, learning rate $l_r=2$, and decays $\rho_R=1/2$, $\rho_S=1/2$, and $\rho_C=1/4$.}
    \vspace{-0.4cm}
    \label{fig:profile_ex}
\end{figure}

These decays can be set to control the impact of a customer satisfaction signal in one RG to $\lambda$-profiles in other RGs and across strata, or even across customers if needed. When signals are rare, it can be beneficial to share these signals \edit{(Steps 3 and 4)}, as customers may have overarching preferences in addition to RG-specific preferences. However, as signals become more common, it may be preferable to set $\rho_S=0$ to prevent sharing of signals across RGs, allowing better convergence of $\lambda$ to the true preference in each RG. An example of the update computation is given in Figure \ref{fig:profile_ex}. The customer has two subscriptions and each has two resource groups. Under each resource group, there is a $\lambda$-profile score for Burstable (B), General Purpose (G), and Memory-Optimized (M) service (see Section~\ref{sec:motivate}). When a signal is triggered for General Purpose database for resource group $r_{21}$, a chain-impact of updates will be made to all the other related services based on the defied decay rate $l_r$. \edit{The same propagation mechanism can be generalized to cross-customer learning where calibration for more propagation rates is required.}

\section{Implementation}
\label{sec:implement}

To apply \sysname to \pg customer data in production, we prioritize low latency and compute price. We therefore perform daily batch inference offline to generate predictions for all customers with existing \pg databases, using the following architecture (see Figure~\ref{fig:scaling}): 
\begin{figure*}[t]
    \includegraphics[width=0.8\textwidth]{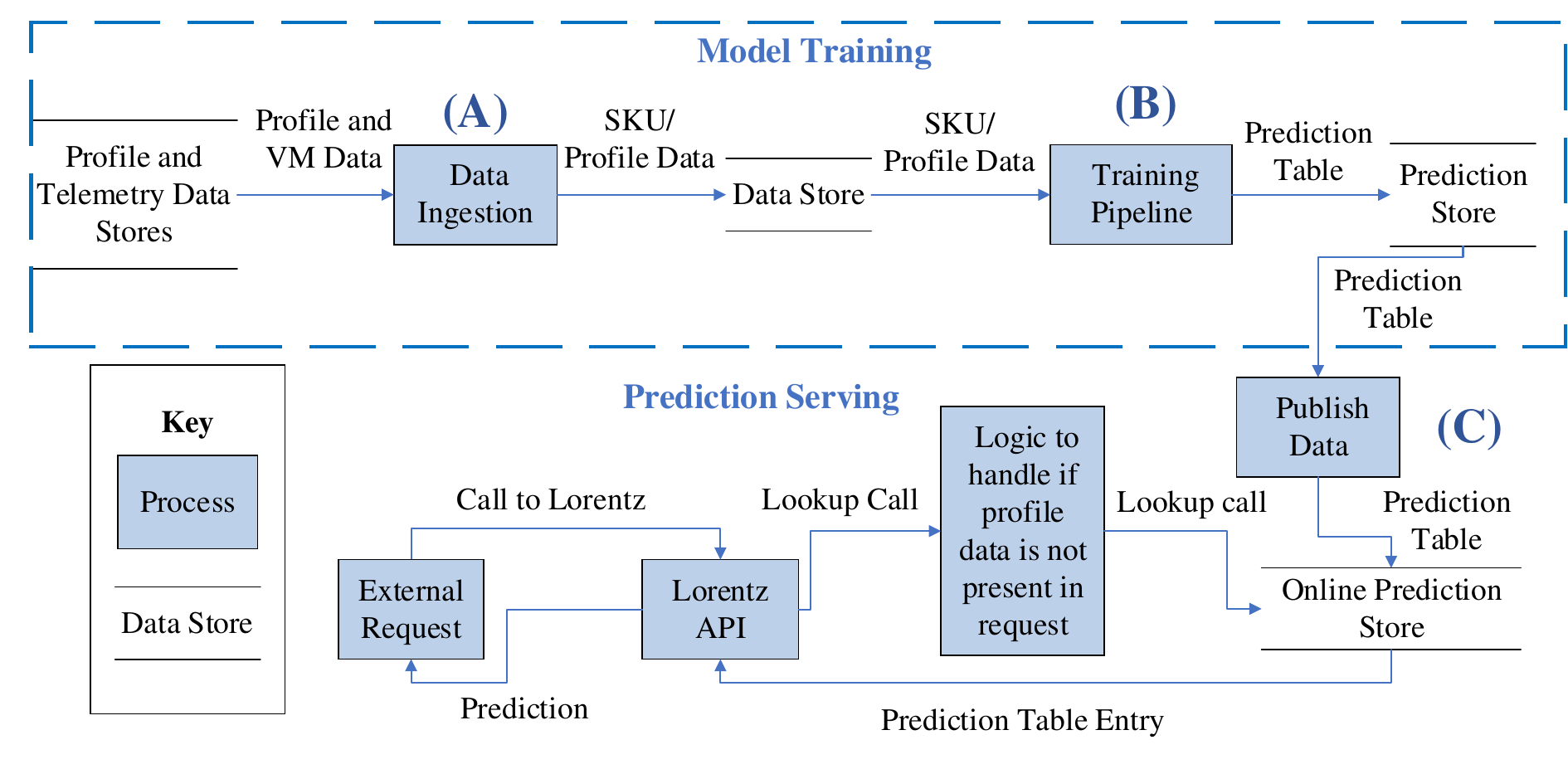}
    \vspace{-0.4cm}
    \caption{\edit{\sysname leverages an offline pipeline for frequent retraining and prediction pre-computation.}}
        \vspace{-0.2cm}
    \label{fig:scaling}
\end{figure*}

\begin{itemize}
\item Data Integration (A): Each batch of \edit{\metadata} and VM usage/health data is ingested via big data processing pipelines from \edit{\metadata} and telemetry data stores onto the compute resource used to train the model.
\item Training Pipeline (B): We use this data as the input into our training and prediction pipeline. The pipeline executes data cleaning and validation, retrains \sysname on the fresh data, and confirms that the new model's performance on a validation dataset is acceptable. The training and prediction pipeline also stores the trained model and its performance metrics for offline experimentation and model improvement. We then use the retrained \sysname model to make predictions in batch, computing an SKU recommendation for each unique [hierarchy level, feature value, server offering] key represented in our dataset.
\item Publish Data (C): We use a cloud-based ETL service to copy the new predictions to an online prediction store which provides low-latency lookup with built-in authentication and data versioning.
\end{itemize}

For inference, \sysname receives SKU prediction requests from downstream services for all \pg customers; each request contains the proposed instance's \edit{feature} vector $\mathbf{x}$. From the online prediction store, we return the prediction corresponding to the most granular hierarchy level in $\mathbf{x}$ whose value is present in the store; if no values in $\mathbf{x}$ are present, we return a default prediction.

\edit{The final output presented to the user includes not only the recommended SKU but also the full rationale behind the choice for explainability. This includes information about both the DB instances identified as "similar" to this user's new request and those instances' chosen SKUs. We also provide the user's current value of cost-performance sensitivity score $\lambda$ (i.e., their perceived preference on the performance-cost trade-off), allowing them to adjust this value to their liking.}

\subsubsection*{\textbf{Alternate implementations}}
It's important to address that in this design we prioritize low latency and compute price over other potentially desirable attributes requiring different architecture approaches. For example, if cloud operators instead prioritized prediction freshness or robustness to newly-created customer accounts, they could implement \sysname using an online architecture where the model is served via an inference endpoint, rather than pre-computing predictions on a schedule.

\section{Experiments}
\label{sec:experiments}
\begin{figure}[htbp]
    \centering
    \includegraphics[width=0.5\linewidth]{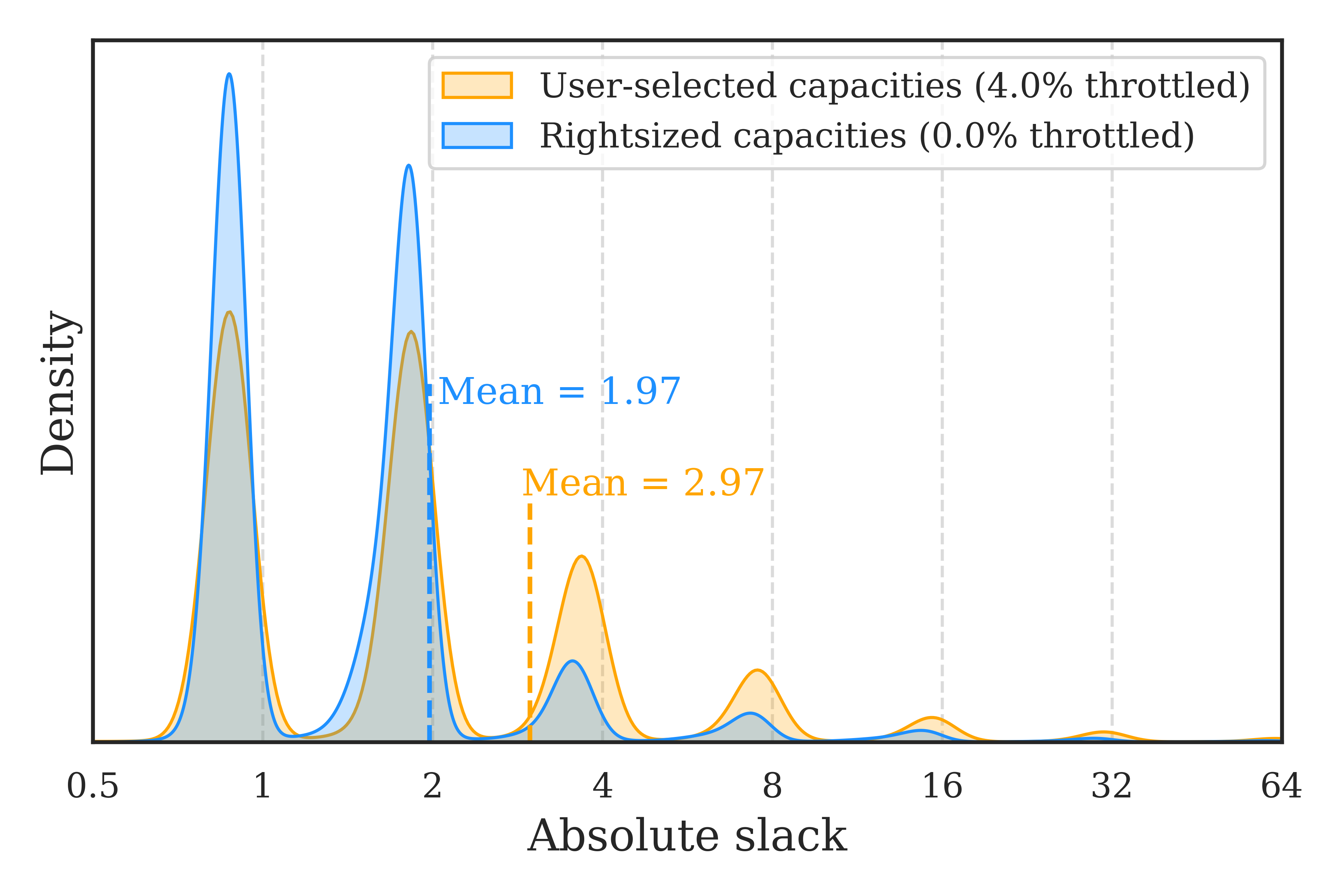} 
    \vspace{-0.1cm}
    \caption{Capacity rightsizing reduces slack and throttling over user selections \edit{based on empirical data}.}
    \label{fig:stage1_plot}
\end{figure}
This section describes the experiments we use to evaluate each stage of \sysname.
Table \ref{tab:hyperparams} describes the hyperparameters used for each stage's experiments.
\begin{table}[htbp]
    \centering
    \caption{Hyperparameters}
    \vspace{-0.3cm}
    \resizebox{7.5cm}{!}{
    \vspace{-0.6cm}
    \begin{tabular}{lll}
        \toprule
        \textbf{Model} & \textbf{Hyperparameters} \\
        \midrule
        Stage 1: Rightsizer & $T=5$ min, $\eta=0.95$, $s_{\text{CPU}}^*=0.5$, \\
        &$\tau=0$, $K=1$\\
        \midrule
        Stage 2: Capacity recommenders & Train/val/test=80/10/10 \\
        Hierarchical model &$p=50$, $\gamma = 0.6$\\
        Target encoder & \# trees = 100, $\xi=$log$_2$ \\
        \midrule
        Stage 3: Personalizer & learning rate = 0.3,\\
        & signal decay = 0.25\\
        \bottomrule
    \end{tabular}
    }
    \label{tab:hyperparams}
    \vspace{-0.3cm}
\end{table}

\edit{In the hierarchical model, $p=50^{\text{th}}$ percentile serves as a balanced, outlier-robust choice. The hierarchy threshold $\gamma$ is empirically selected to include only the observed group of strong hierarchies, and $\xi=$log$_2$ maps the capacity options to linear space.}

\subsection{Stage 1: Rightsizing}

The goal of the rightsizing stage is to compute new resource capacities that minimize slack and throttling for existing workloads given that many have been over/under-provisioned. In our \pg dataset, we find significant disparities between user-selected and rightsized capacities; see Section 2.2 and Figures \ref{fig:pie-chart} \& \ref{fig:skus} for a complete summary.

We further evaluate the effectiveness of this stage on the \pg workloads $W$ by comparing the observed slack and throttling between user-selected capacities and our rightsized capacities, computed extremely conservatively using $s^*=0.5$, i.e., 50\% CPU utilization, and $\tau=0$, i.e., 0\% throttling limit as in Equation~\eqref{eq:rightsizing}. Here, we use similar definitions for slack and throttling as in Stage 1. In Figure~\ref{fig:stage1_plot}, we show the distribution of \textit{absolute slack}, i.e., $S_{w,r}(c_r) \cdot c$, to better align with the business goal of minimizing the global resource volume provisioned, and compute each capacity set's \textit{throttling ratio} as the ratio of workloads $\textbf{w} \in W$ with $T_w(\mathbf{c}) > \tau$. That absolute slack weights higher-capacity servers higher is desirable in our business context, since resource costs generally scale linearly with capacity.

Figure \ref{fig:stage1_plot} shows that our rightsized capacities eliminate throttling entirely, while still reducing slack by 34\%. Note that absolute slack distributions are modal around powers of 2 because the candidate capacities are, for the most part, powers of 2. 

The challenge, of course, is predicting these rightsized capacity labels before VMs are created; the following section describes \sysname's solution to this task.

\begin{figure}[htbp]
    \centering
    \includegraphics[width=.5\linewidth]{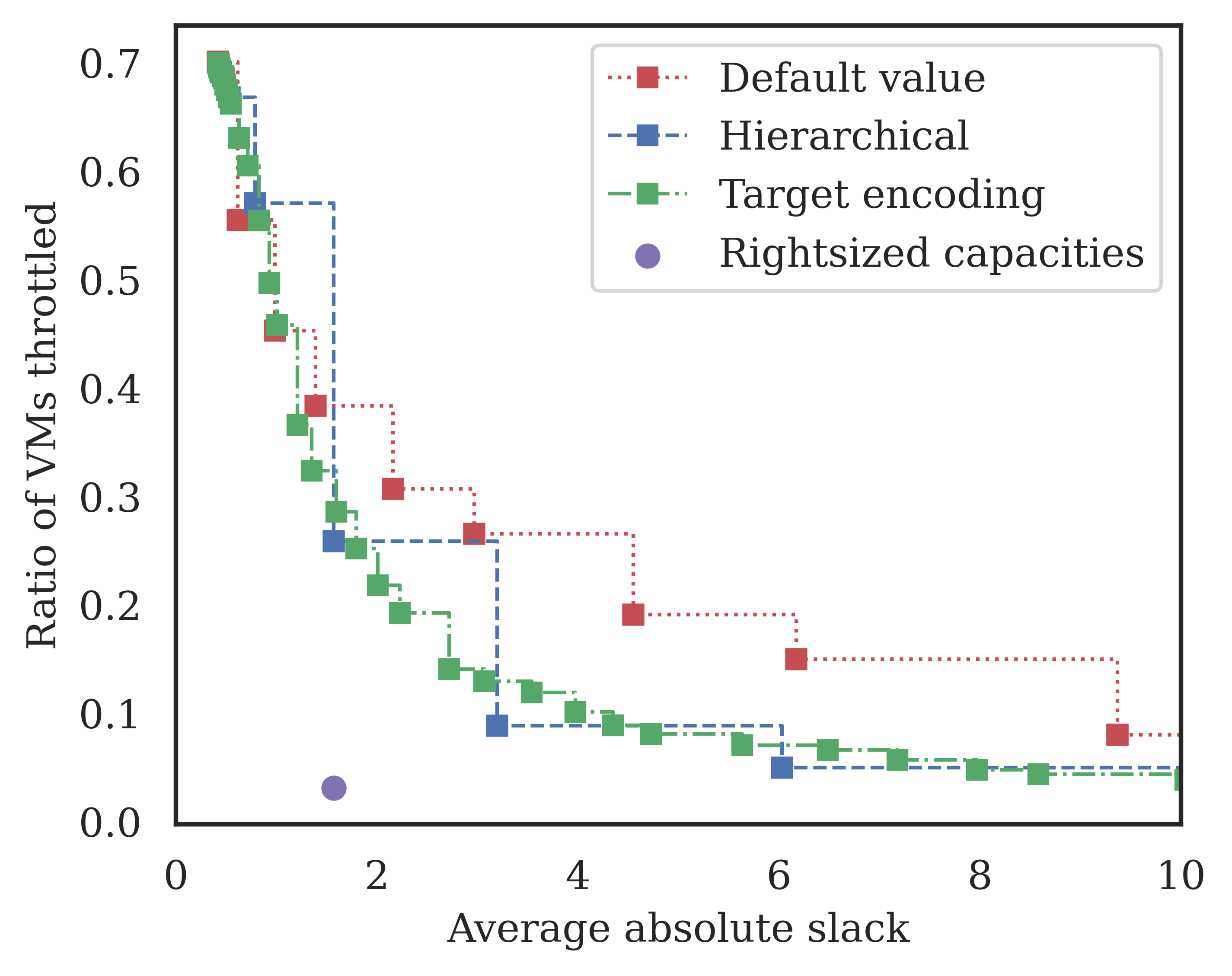} 
        \vspace{-0.3cm}
    \caption{\sysname provisioners improve the slack/throttling Pareto frontier over baselines.}
    \label{fig:stage2_pareto}
\end{figure}
\begin{figure}[htbp]
    \centering
    \includegraphics[width=0.5\linewidth]{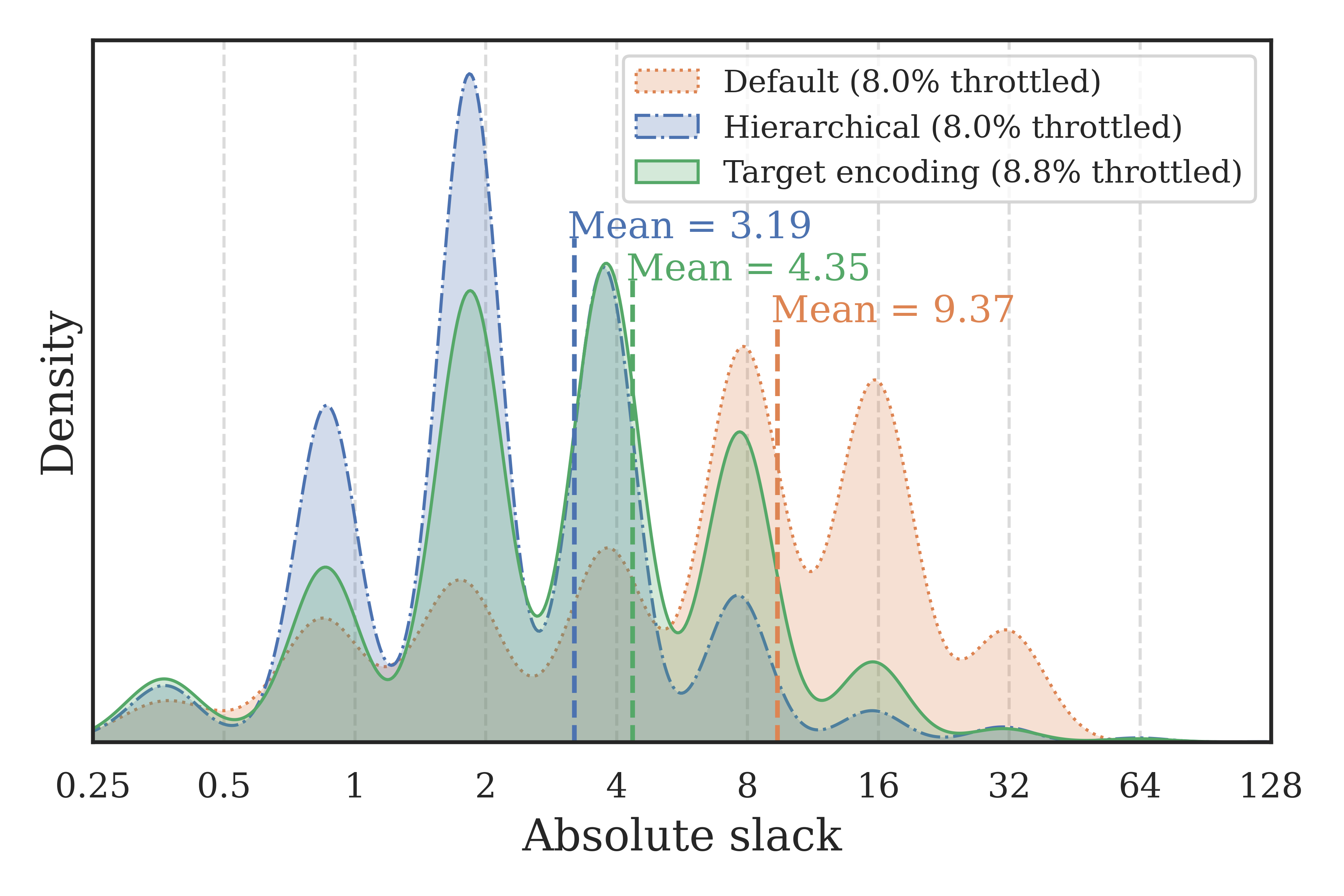} 
        \vspace{-0.3cm}
    \caption{\sysname provisioners improve slack while keeping average throttling<10\% over baseline. 
    }
    \label{fig:stage2_kde}
\end{figure}

\begin{figure}[htbp]
    \centering
    \includegraphics[width=.5\linewidth]{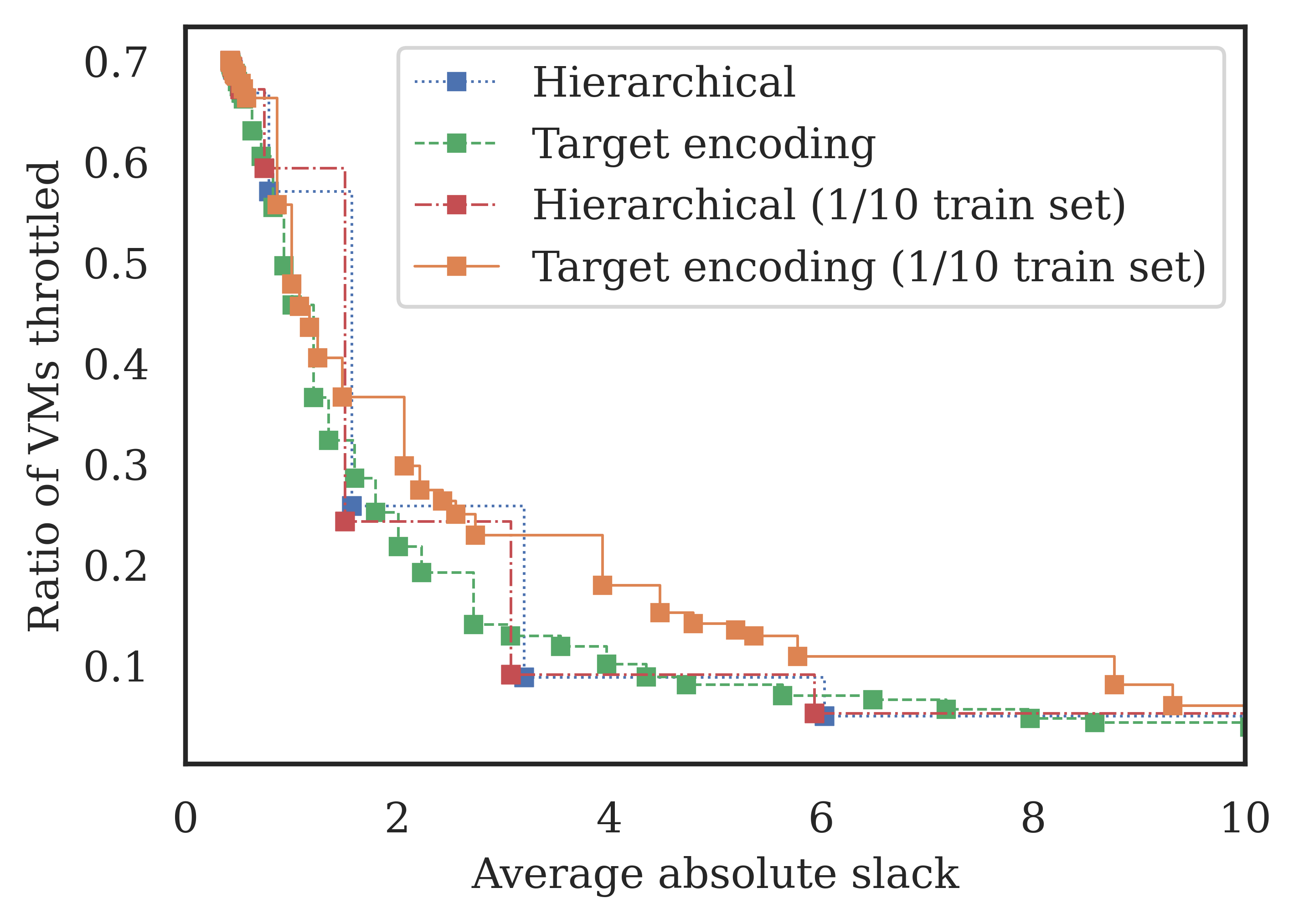} 
        \vspace{-0.3cm}
    \caption{The hierarchical provisioner is robust to data quantity.}
    \label{fig:stage2_pareto_small}
\end{figure}

\subsection{Evaluating \edit{end-to-end provisioners}}


In this section, we evaluate the ability of \sysname provisioners to recover the rightsized capacity labels \edit{following Stages 1 and 2}, given useful features (computed only based on \edit{\metadata}) and diverse workloads. The dataset of \pg resources is extremely left-skewed, containing many low-usage workloads: the mean maximum utilization is only 1.2 vCores. Further, our rightsizing algorithm selects the minimum capacity choice for 86\% of DBs, and one of the two smallest choices 95\% of the time. As a result, our provisioners achieve similar average slack to default-value baselines at most throttling ratios, with a maximum possible slack reduction of 33\%. \edit{We also evaluated the provisioner models based on their aggregate vCores provisioned and hours throttled, extrapolated from the test set to a count of ~67k servers (the total number of one DB offer for \pg), achieving 27\% (Hierarchical) and 8\% (Target Encoding) reduction in cost compared to user selection.}

\subsubsection*{\textbf{Synthetic workload upscaling.}}  
To better demonstrate the effectiveness of our provisioner models, we therefore generate synthetic usage telemetry, increasing the diversity of our label set by scaling up some of the workloads according to their \edit{\metadata}. 

To upscale the \pg workloads, we first compute the hierarchy present in their \edit{profile} features using Step 1 of the hierarchical provisioner, from most to least granular.
We then upscale workloads with the following steps:
(1) Select three features throughout the hierarchy, and assign them global scale factors--ResourceGroup: 1, CloudCustomerGuid: 1, VerticalName: 3;
(2) Per feature, assign each unique value a scale factor: either that feature's global scale factor or 0, with equal likelihood;
(3) Compute each workload's total scale factor $\chi_w$ by applying its feature values' assigned scale factors in combination, producing a result between 0 and $1+1+3=5$;
(4) Upscale each workload as $2^{\chi_w}\cdot w[n]$;
(5) Recompute the rightsized capacities.
After upscaling, the dataset is much more diverse: the mean maximum utilization is 9.0 vCores and the rightsized capacity is one of the lowest two choices for only 55\% of workloads. 

\subsubsection*{\textbf{Provisioner evaluation.}} We train the provisioner models on a 80\%/10\%/10\% train/validation/test split using the hyperparameters described in Table \ref{tab:hyperparams}. As described in Section \ref{sec:dataset}, we fit distinct models to each stratified server offering; each of the stratified models uses the same hyperparameters, and all results are averaged across the offerings.

We evaluate the provisioner models on the upscaled workloads by computing their absolute slack and throttling ratio, as for Stage 1. Since users have variable sensitivities to slack and throttling, we cannot produce a single metric defining model performance; instead, we observe the Pareto frontiers of our models. We compute the Pareto curve for each model to evaluate the optimality of the recommended SKUs.
For the hierarchical and target encoding provisioners, we build the Pareto curves by scaling all recommendations up and down by varying powers of two.
Each scale factor produces one Pareto curve point: for example, scaling all predictions by $2^{-2.5}$ produces a low-slack, high-throttling point.
We construct our default-value baseline by evaluating different sets of default values, where each set defines one default value for each of the three server offerings (e.g., Burstable, General-Purpose, MemoryOptimized). 
Note that since many users (89\%) in the original observation select either the default vCore value or the next larger option, we consider this default baseline a reasonable approximation of user behavior after the scaling. 



Figure \ref{fig:stage2_pareto} shows that the Pareto frontiers of both provisioner models notably exceed those of the baselines along both the slack and throttling axes using the synthetic workload. The hierarchical model performs particularly well, although its Pareto curve is coarser since its predictions are discrete. On the other hand, the target encoder exchanges a small retraction in performance for incredible flexibility to be tuned along the slack/throttling spectrum. \edit{Target encoder yields detailed Pareto curves with ample data, while the hierarchical model delivers more robust performance in situations with limited data. For both methods, relying solely on \edit{\metadata} doesn't reach a perfect solution due to the variation of SKU choices within the same bucket. In this paper, we aimed to propose a general framework that can be extended with the availability of more data, potentially reducing the variability within each bucket.}

\edit{We compare the single point on each model's Pareto curve as in Figure~\ref{fig:stage2_pareto} that minimizes slack with a throttling ratio < 10\%,} a middle ground in the slack/throttling tradeoff. The probability density function plot in Figure \ref{fig:stage2_kde} describes the models' slack distributions at these points; we again observe here that both provisioner models outperform the baselines. While achieving a similar level of throttling, the hierarchical provisioner reduces mean slack by 66\%, while the target encoder provides a 54\% reduction.
\edit{We observe that workloads with greater variation are more throttling-prone, where cost improvements outweigh perfformance degradation. \sysname also performs consistently across all capacity needs, skewed towards under-provisioning for large outliers (>32 vCores).}

\subsubsection*{\textbf{Smaller training set.}} To evaluate the provisioners' robustness to data quantity, we provide in Figure \ref{fig:stage2_pareto_small} their performance on a training set containing 10\% of the samples in the full training set, randomly sampled, after the workload upscaling. We find that although the performance of the target encoder suffers with fewer training samples, the hierarchical model yields nearly equivalent performance despite the drastically smaller training set. Those deploying \sysname should therefore select the most appropriate provisioner based on both the size of their training set and the flexibility required for their use case: if data is plentiful, take advantage of the target encoder's very granular Pareto curve, while if data is scarce, the hierarchical model will offer more robust performance.

The \sysname provisioner models are effective despite being lightweight, and require relatively few features and parameters. This makes them especially feasible for deployment to realize even minor improvements over baselines. As a result, though, it's important that any VM \edit{\metadata} used as inputs is verified to be informative of their corresponding workloads.

\subsection{Evaluating Stage 3: Personalization}
While the evaluation of the personalization requires labled data (true sensitivity scores) and a longer time frame of observation for feedbacks, in order to test the convergence of our algorithm, we simulate the behavior of three customers assuming a ground truth of cost/performance sensitivity: Alice ($\lambda_A=0$), Bob ($\lambda_B=1.5$), and Charlie ($\lambda_C=-1.5$). Each of these customers has three subscriptions with their own associated cost/performance sensitivity, \edit{``Dev" ($\lambda_{\text{dev}}=-1$), ``Prod1" ($\lambda_{\text{prod1}}=0.5$), and ``Prod2"} ($\lambda_{\text{prod2}}=1.5$). Thus, the true cost/performance sensitivity for Bob's Prod2 subscription is $\lambda_{B,\text{prod2}}=\lambda_B+\lambda_{\text{prod2}}=3$, and so on.
Each subscription has three resource groups, and within each resource group, we randomly provision $n_{R}\in\{1,...,5\}$ resources.
For simplicity, we draw a random $\mathbf{c}^*$ from the set of $\mathcal{C}$=$\{1,2,4,...,128\}$ representing the recommendation from Stage 2 for each resource. We assume that the error of such generated recommendation follows a log-normal distribution, $\log_2(\epsilon)\sim N(0,\sigma^2)$, and the optimal candidate (before customization) should be: $\Bar{\mathbf{c}}^* = \mathbf{c}^* +\epsilon$.
For each resource, considering the customer preference, the optimal capacity that the corresponding customer will be most satisfied with can be calculated as: 
\begin{align}
  \mathbf{\Bar{c}}^{**}=2^{\lambda_{\{\text{A,B,C}\},\{\text{dev,prod1,prod2}\}}} \Bar{\mathbf{c}}^*  
\end{align}
For each resource, we initialize the personalization parameters $\widehat{\lambda}_0 = 0$, making the initial SKU recommendation $\mathbf{c}^{**}_0=2^{\widehat{\lambda}_0}\mathbf{c}^* = \mathbf{c}^*$.

We then simulate the profile learning from this initial data set in a three-step loop: (Step 1) Generate signals, (Step 2) Update profiles, (Step 3) Recompute predictions.

\paragraph{Step 1: Generate signals.} Ideally, at time $t$, each customer gives us a signal for each poorly-provisioned resource: if the resource is over-provisioned ($\mathbf{c}^{**}_t>\mathbf{\Bar{c}}^{**}$), we get a $-1$ signal, and if the resource is under-provisioned ($\mathbf{c}^{**}_t<\mathbf{\Bar{c}}^{**}$), we get a $+1$ signal. We introduce two more parameters: signal rate (the probability that a signal is generated at all, simulating our ability to collect signals from customers) and signal noise (the probability that we collect an \textit{incorrect} signal, flipping the sign). These two controls allow us to confirm the convergence of the profiles over time when signal collection is imperfect.

\paragraph{Step 2: Update Profiles.} After signal generation, we update the stored profiles for each resource group using the algorithm outlined in Section \ref{sec:personalization}. 

\paragraph{Step 3: Recompute Predictions.} 
The predictions at time $t>0$ are given by $\mathbf{c}^{**}_t=2^{\widehat{\lambda}_i}\mathbf{c}^{*}$, discretized to $\mathcal{C}$. 
The difference between the personalized recommendation and the optimal capacity is: 
\begin{align}
    \Delta &= \mathbf{c}^{**}_t - \mathbf{\Bar{c}}^{**} \nonumber\\ \nonumber
    &=2^{\widehat{\lambda}_t}\mathbf{c}^* - 2^{\lambda_{\{\text{A,B,C}\},\{\text{dev,prod1,prod2}\}}} \Bar{\mathbf{c}}^*  \\\nonumber
    &=2^{\widehat{\lambda}_t}\mathbf{c}^* - 2^{\lambda_{\{\text{A,B,C}\},\{\text{dev,prod1,prod2}\}}} (\mathbf{c}^* +\epsilon) \\
    &=(2^{\widehat{\lambda}_t} - 2^{\lambda_{\{\text{A,B,C}\},\{\text{dev,prod1,prod2}\}}}) \mathbf{c}^* - M\epsilon,
\end{align}
where $M$ is a constant.
When $|\lambda_{\{\text{A,B,C}\},\{\text{dev,prod1,prod2}\}}-\widehat{\lambda}_i|$ is small enough, the recommended capacity will be close to the optimal while only the error of Stage 2's prediction dominates.

\subsubsection{Results}

\begin{figure}
    \centering
    \includegraphics[width=0.5\linewidth]{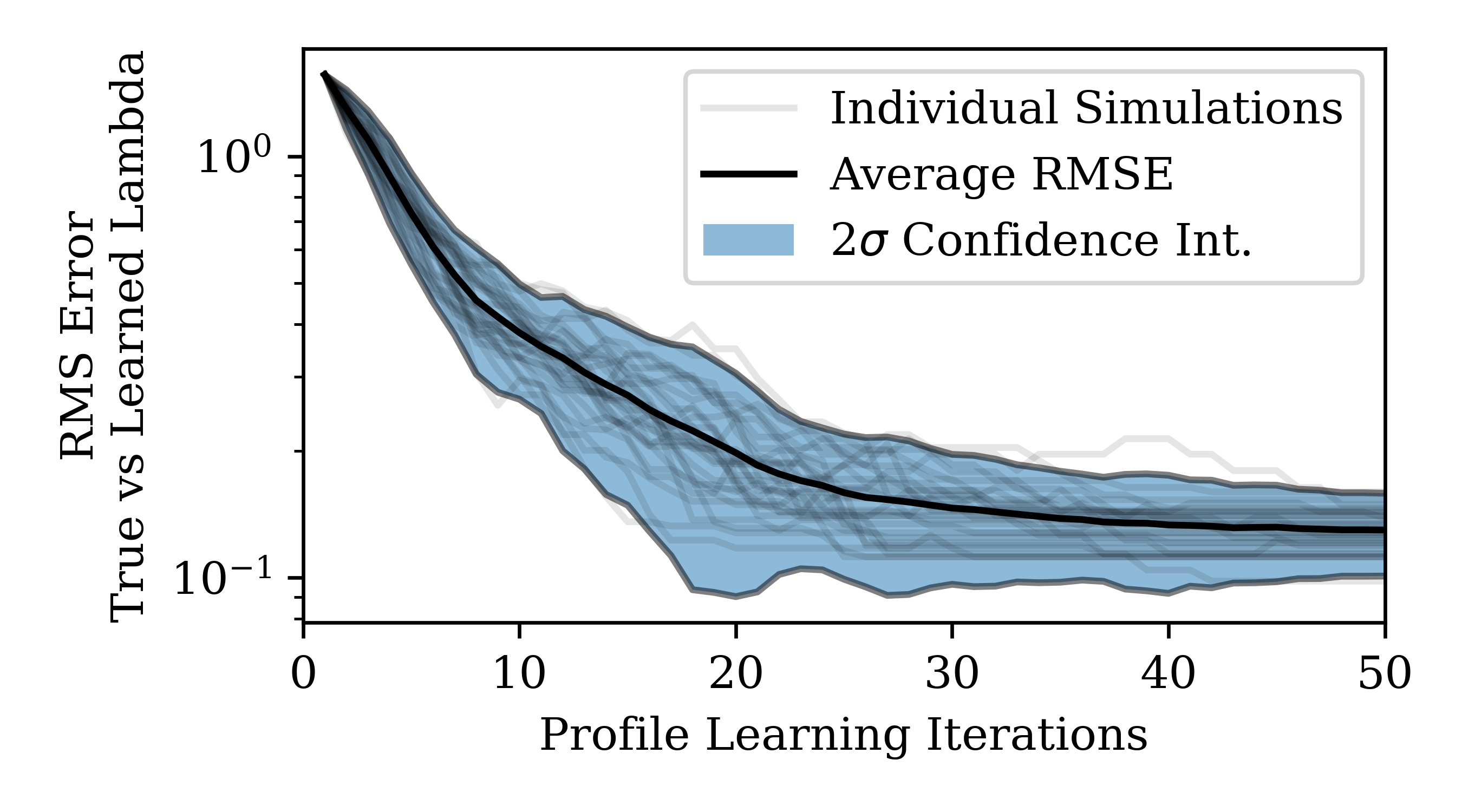}
        \vspace{-0.4cm}
    \caption{\sysname's Stage 3 converges rapidly in all simulations.}
    \label{fig:one_setting_sims}
    \vspace{-0.5cm}
\end{figure}

We run two main experiments using the above settings.
First, we demonstrate that the Stage 3 algorithm converges reliably for a given setting. We set the signal rate to 40\% (40\% of misconfigured systems generate a signal at each iteration), the signal noise to 13\% (13\% of geenrated signals are multiplied by $-1$), the Stage 2 error $\sigma=0.1$. In Figure \ref{fig:one_setting_sims}, we plot individual simulations (gray lines), the average simulation (black line), and an estimated point-wise 95\% confidence interval (2 standard errors). On average, customer profiles reach a root mean squared error (RMSE) of 0.15 within 30 iterations. Profile learning ceases once each customer's system is accurately provisioned.

\begin{figure}
    \centering
    \includegraphics[width=0.5\linewidth]{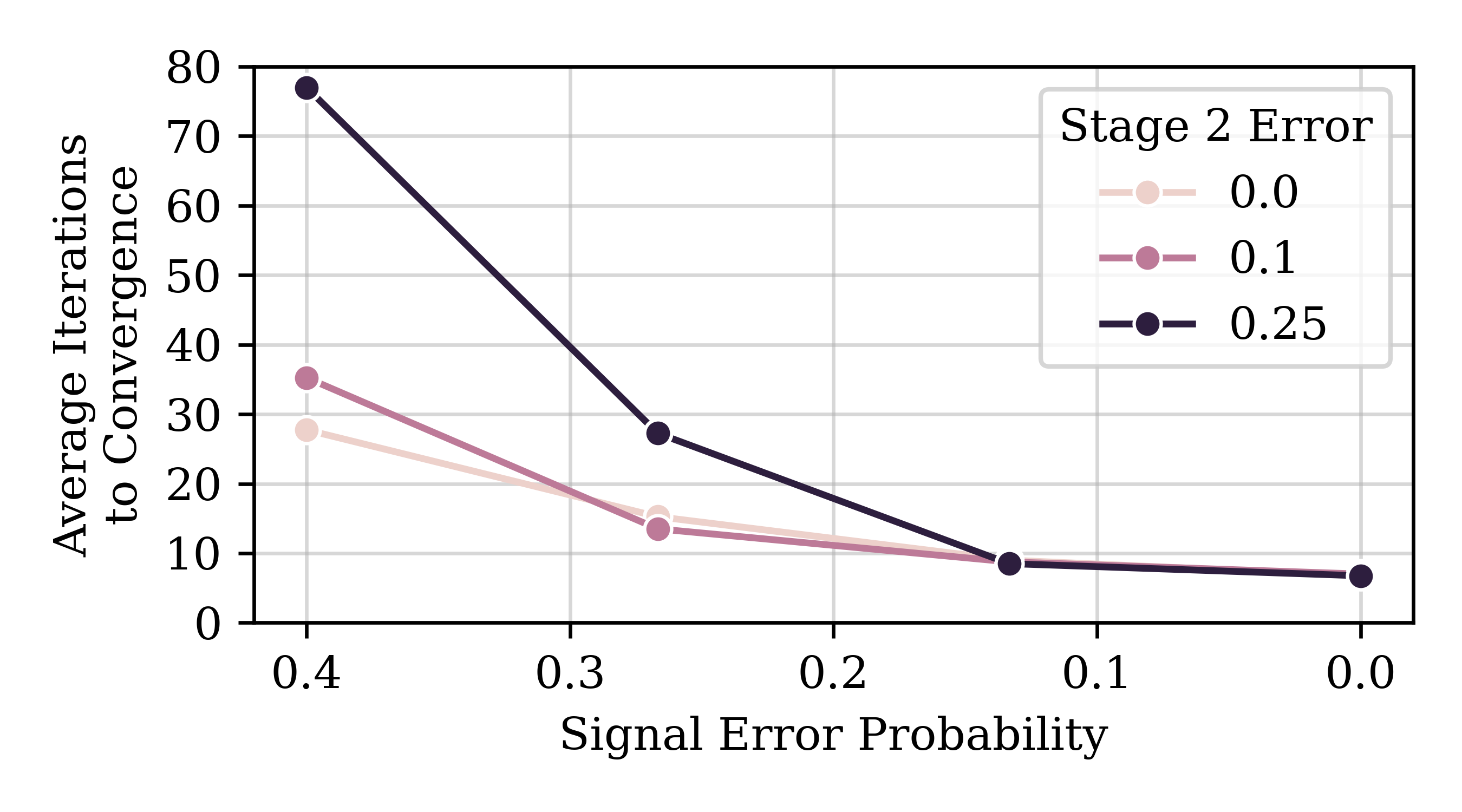}
        \vspace{-0.4cm}
    \caption{Stage 3 converges rapidly when customer signals are accurate.}
    \label{fig:multi_setting_sims}
\end{figure}

Second, we evaluate the personalizer at each element in the Cartesian product of signal error values $\{0\%, 13\%, 26\%, 40\%\}$ and Stage 2 error ($\sigma$) values $\{0.0, 0.1, 0.25\}$. At each element, we run multiple simulations at signal rates of $\{10\%, 40\%, 70\%, 100\%\}$ respectively, averaging the convergence times over these experiments to produce one point in Figure~\ref{fig:multi_setting_sims}. We define ``convergence'' here as the first iteration that the 80$^\text{th}$ percentile of customer profiles have errors below 0.5, or $|\lambda_*-\widehat{\lambda}|\le0.5$. At this point, 80\% of customers are receiving recommendations within half a step of their true preference, typically rounding to their preferred capacity. In general, these results indicate that (i) \sysname rapidly personalizes results to individual customers if signals can be accurately classified and Stage 2 results are accurate, and (ii) accurate cost/performance signal classification can overcome both sparse or rare signals and inaccurate Stage 2 predictions.

\section{Related Work}
\label{sec:related}

SKU selection has perpetually posed a challenge. Prior studies have delved into situations involving workload migration, tapping into existing traces to aid in decision-making. Examples are Cloud Target Selection (CTS)~\cite{kopaneli2015model} and Cloudle~\cite{kang2010cloudle} where systematic processes on the foundation of exhaustive questionnaires are created, enabling users to outline their requirements. However, these tools primarily aimed at streamlining the decision-making process but were not able to fully automate the process. 
REMICS~\cite{sadovykh2011remics} undertook an in-depth exploration of prevailing workloads, encompassing source code, configuration files, and execution traces. This culminated in the development of a model-driven tool to facilitate migration. A recent addition to this domain, \doppler~\cite{cahoon2022doppler}, delved into harnessing machine learning to assess the advantages of selecting optimal SKUs. This approach was grounded in historical resource consumption and insights gleaned from existing customers, factoring in preferences for price-performance trade-offs. However, these methodologies all share a common trait--the need for extensive input regarding the workload, rendering them impractical for our particular scenario.


Other related lines of research revolve around resource scaling after provisioning, encompassing both reactive scaling~\cite{das2016automated,delimitrou2014quasar,floratou2017dhalion} and proactive scaling~\cite{poppe2022moneyball, poppe2020seagull, gong2010press, islam2012empirical, cortez2017resource}. Techniques such as machine learning~\cite{poppe2022moneyball, poppe2020seagull,calheiros2014workload,roy2011efficient,islam2012empirical,khan2012workload,taft2018p}, signal processing~\cite{gong2010press}, and statistical methods~\cite{islam2012empirical} are employed to predict future usage while accounting for performance constraints such as throttling impact and QoS requirements~\cite{padala2009automated,delimitrou2014quasar}, aiding in the determination of optimal resource allocation.











\section{Conclusion}
\label{sec:conclusion}





In this study, we introduced \sysname, a system to generate SKU recommendations for \textit{newly-provisioned} cloud resources without relying on telemetry data or workload traces. \sysname employs a three-stage approach: capacity rightsizing, capacity recommendation, and personalization, allowing it to accommodate diverse user preferences for price and performance across various services. Through a continuous learning pipeline, the system dynamically adjusts preference scores based on user feedback signals, such as Customer Reported Incident (CRI) information. Evaluation on production data demonstrated that \sysname eliminates >60\% of wasted COGS compared to our baseline, indicating its effectiveness in optimizing cloud resource allocation. 

\sysname can be extended to suggest capacities for multiple resources, offering comprehensive SKU recommendations. Additionally, incorporating more entries of \edit{\metadata} features could potentially enhance the recommendation accuracy as well as enable recommendations of suitable server offerings among different types, adding substantial value. 


\balance
\bibliographystyle{ACM-Reference-Format}
\bibliography{sample}

\end{document}